\documentclass[preprintnumbers,amsmath,amssymb,floatfix,11pt,prd,onecolumn,superscriptaddress,nofootinbib]{revtex4}
\usepackage{diagbox}
\usepackage{latexsym}
\usepackage{epsfig}
\usepackage{epstopdf,float}
\usepackage{graphicx}
\usepackage{amssymb}
\usepackage{amsmath}
\usepackage{dcolumn}
\usepackage{bm}
\usepackage{color}
\usepackage{comment}
\usepackage[shortlabels]{enumitem}
\usepackage{subfigure}
\usepackage{multirow}
\usepackage{xcolor}
\begin{document}
\title{\bf Imprints of Dark Matter on the Shadow and Polarization Images of a Black Hole Illuminated by Various Thick Disks}
\author{Muhammad Israr Aslam}
\altaffiliation{mrisraraslam@gmail.com,
israr.aslam@umt.edu.pk}\affiliation{Department of Mathematics,
School of Science, University of Management and Technology,
Lahore-$54770$, Pakistan.}
\author{Rabia Saleem}
\altaffiliation{rabiasaleem@cuilahore.edu.pk}\affiliation{Department
of Mathematics, COMSATS University Islamabad, Lahore Campus,
Lahore-$54000$ Pakistan.}
\author{Chen-Yu Yang}
\altaffiliation{chenyu\_yang2024@163.com}\affiliation{Department of
Mechanics, Chongqing Jiaotong University, Chongqing $400000$,
People's Republic of China}
\author{Xiao-Xiong Zeng}
\altaffiliation{Corresponding author:
xxzengphysics@163.com}\affiliation{College of Physics and Electronic
Engineering, Chongqing Normal University, Chongqing $401331$,
People's Republic of China}

\begin{abstract}
Based on two distinct thick accretion flow disk models, such as a
phenomenological RIAF-like model and an analytical Hou disk model,
we investigate the impact of relevant parameters on the visual
characteristics of the Schwarzschild black hole (BH) surrounded by perfect fluid dark matter (PFDM). We impose a general relativistic radiative transfer equation to
determine the synchrotron emission from thermal electrons and
generate horizon-scale images. In the RIAF-like model, we notice
that the corresponding photon ring and central dark region are
expanded with the aid of the PFDM parameter $\eta$, with brightness
asymmetries originating at higher inclination angles and closely
tied to flow dynamics and emission anisotropy. The fundamental
difference between isotropic and anisotropic radiation is that
anisotropy introduces vertical distortions in the higher-order
images, resulting in an elliptical appearance. For the Hou disk
model, the observed images produce narrower rings and dark
interiors, while polarization patterns trace the brightness
distribution and changes with the variations of the inclination angle
and PFDM parameter $\eta$, which reflects the spacetime signature.
All these results indicate that the observed intensity and
polarization characteristics in the framework of thick disk models
may serve as valuable probes of underlying spacetime geometry and
the accretion-dynamics close to the horizon.
\end{abstract}
\date{\today}
\maketitle

\section{Introduction}
Einstein's theory of general relativity (GR) stands as one of
the most remarkable scientific achievements of the last century. It
has successfully addressed a large range of complex phenomena, from
the dynamics within our solar system to the cosmological scale
structure of the universe, all supported by strong observational
evidence. It has led to various fascinating discoveries associated
with gravity and has introduced profound advancements in
understanding various astrophysical phenomena within the fabric of
modern cosmology. Black holes are one of the fundamental solutions
to Einstein's equations, known as cast-iron exceptional predictions
of GR \cite{p1}. Nowadays, BHs are known as the most intriguing
astronomical objects possessing the appealing attributes of strong
gravitational fields. The strong gravitational field around BH
prevents anything from emitting, causing any matter or radiation
interacting within its vicinity to be inevitably absorbed. The
recent images of the super-massive BHs M$87^{\ast}$ and Sgr
$A^{\ast}$, which was released by the Event Horizon Telescope (EHT)
have opened a new window in the theoretical as well as observational
studies of BH physics \cite{sd3,sd4,sd6,sd10,jp20,jp21}. These
groundbreaking discoveries not only validate the theoretical
predictions of GR but also provide valuable insights into the
complex characteristics of BHs and the nature of extreme cosmic
environments, thereby ushering in a new path of BH astronomy. An
astrophysical BH preserves stable space-time geometry but can be
illuminated by surrounding luminous accretion material, producing a
wide variety of observable shapes and colors. When light from the
accretion flow approaches the BH, its strong gravitational field
bends the light toward the singularity, enabling the study of the
optical signatures produced by the accretion. The intricate
structure of the BH shadow is determined by the underlying dynamics
of photon orbits and the surrounding space-time geometry.
Observations from the EHT have revealed a central dark region,
referred to as the BH shadow, encircled by a compact and asymmetric
bright ring, commonly known as the photon ring.

The study of BH shadows dates back to the early developments of GR.
In the 1960s, the theoretical framework describing the shadow of a
Schwarzschild BH was first proposed in \cite{sd11}, where a formula
was derived to determine the angular radius of the BH shadow.
Subsequently, Bardeen investigated the shadow of a Kerr BH and,
utilizing the separability of the null geodesic equations, developed
a formalism to determine the boundary of the shadow \cite{sd12}.
Despite these important developments, the BH shadow was long
regarded as a purely theoretical concept, seemingly beyond
experimental observation. In \cite{sd13}, however, the possibility
of directly observing the BH shadow at the center of our Milky Way
galaxy was proposed, along with the necessary observational
requirements. The study of the visual features of BHs serves as a
crucial key to uncovering their fundamental nature. With the
successful observational imaging of BHs, extensive theoretical
research has been actively pursued in recent years along with
realistic mathematical frameworks. Inspired by pioneer works,
scientists have studied the observational characteristics of
super-massive BHs through the effective implementation of numerical
simulations. Beyond the fundamental BH solutions such as
Schwarzschild and Kerr BHs, numerous complicated BH models have
been investigated to analyze their visual appearances (see Refs.
\cite{pol1,pol2,pol3,pol4}) and so on. Consequently, considering the
framework of a four-dimensional Gauss-Bonnet BH, the authors in
\cite{p24} discussed the visual characteristics of BH shadows with
different accretions of matter. Parbin et al. \cite{pol5} discussed
the influence of the axionic parameter on the shadow of slowly rotating
BH in Chern-Simons modified gravity. Gralla et al. \cite{pol6}
investigated the shadows and photon ring images of the Schwarzschild BH
within the mechanism of geometrically thin disk accretion flow
model. Moreover, much significant research has focused on the
impact of the different accretions of matter on the shadows of the
BH \cite{gkref26,gkref27,sd37,sd38,zeng4}. Consequently, several
significant papers have been devoted to the study of BH shadows with
the help of holographic Einstein ring
\cite{sd21,sd22,sd23,israr1,israr22}, as well as the shadow of
Kerr-like BHs with celestial light sphere and thin accretion disk
\cite{sd34,sd35,sd36,pol7} to mention a few.

In $2021$, the observational team of EHT unveiled the first
polarized images of M$87^{\ast}$ and Sgr $A^{\ast}$
\cite{jp21,pol8,pol9}, revealing a remarkable spiral alignment of
electric field vectors across the emission ring. In the context of
BH astrophysics, polarization in radiation serves as a crucial probe
of magnetic field distributions and the behavior of the surrounding
matter, which can be extracted from BH images by solving the null
geodesic equations governing photon trajectories in the BH
space-time. Additionally, the parallel transport equations must be
solved to trace the transformation of polarization vectors along
photons trajectories. Recently, some authors \cite{pol10,pol11}
utilized a simplified model to investigate polarized images
produced by synchrotron emission from axi-symmetric fluids revolving
around BHs under different magnetic field configurations. The
thermal synchrotron radiation from a BH produces a polarized image,
with the polarization intricately shaped by the magnetic field
orientation in the emitting region, the thermal motion of the
radiating gas, the strong gravitational lensing effect of the BH,
and the parallel transport of polarization in curved space-time.
Considering the framework of the Schwarzschild BH, Narayan and his
collaborators \cite{pol10} employed an approximate expression for
null geodesics derived by Beloborodov \cite{pol12} to analyze the
polarization images of a BH surrounded by hot gas. These results
successfully reproduces the electric vector position angles and the
relative polarized intensity, which is observed in the images of the
M$87^{\ast}$. After that Gelles et al. \cite{th33}, developed a
simplified model with an equatorial emission source in the
background of Kerr BH, which produced the corresponding polarization
images, and discussed the geometric influence of BH spin on the photon
parallel transport. Yang et al. \cite{newyang1} discussed the shadow
and polarization images of rotating BHs in Kalb-Ramond Gravity
illuminated by several thick accretion disks. Further, the
polarization images for different BH frameworks and horizonless
ultra-compact objects have also been discussed widely, for comprehensive
review, one can see Refs.
\cite{newyang2,newyang3,newyang4,young1,th34,th35,th36,young2,th37,th38,th39,young3}.

During the last decades, the investigation of BH dynamics in
different modified gravity and matter field backgrounds has reached
a peak position among the scientific community. Astrophysical BHs
exist within extreme cosmic environments, continuously interacting
with surrounding matter and radiation fields; therefore, they cannot
be accurately described as isolated systems \cite{pol13,pol14}. In
the cosmological mechanism, it is widely acknowledged that the
majority of the universe's matter content consists of dark matter,
whose fundamental nature remains unknown \cite{pol15}. Hence, it is
reasonable to expect that the gravitational field of a realistic BH
is affected by the presence of a surrounding dark matter halo.
Although several models aim to describe the relation between BHs and
dark matter, a particularly tractable and phenomenologically appealing framework, is the Perfect Fluid Dark Matter (PFDM) model.
In this scenario, dark matter is treated as an anisotropic perfect
fluid, enabling its consistent incorporation into the BH space-time
without invoking specific particle-level microphysics \cite{pol16}.
Understanding the intricate properties of dark matter is crucial, not
only for describing galaxy formation and the large-scale structure
of the universe, but also for exploring its relationship with dark
energy, which drives the accelerated expansion of the universe. In
this regard, the PFDM model has emerged as a fascinating mechanism
for investigating dark matter, particularly in the vicinity of BHs.
Unlike classical particle-based dark matter models, the PFDM
framework treats dark matter as a continuous, non-viscous fluid
governed by specific equations of state, offering new insight into
how dark matter varies the space-time geometry and the complex
structure of BHs. For instance, the researchers in \cite{pol17}
consider the framework of PFDM and discuss the dark matter
clustering around BHs, which reveals the important deviations in BH
metrics. Simultaneously, considering the framework of the PFDM model,
many significant papers have been devoted to explore various aspects
of BH dynamics, including gravitational lensing analyses
\cite{pol18}, BH shadows \cite{pol19,pol20}, deflection angles
\cite{pol21}, the super-radiant phenomenon of the massive scalar field
around BHs \cite{supr1} as well as the thermodynamic properties of
BH horizons \cite{pol22,pol23}. Overall, the PFDM model offers a
promising and versatile alternative to conventional dark matter
theories, particularly in explaining how dark matter affects BH
space-times and associated observables across galactic and
cosmological scales.

Motivated by these groundbreaking studies, in this paper, we will
investigate the BH shadows and the polarization images of
Schwarzschild BH in the presence of PFDM configurations. Our main
objective is to explore the impact of model parameters on BH
shadows and the polarization images, as well as the correlation
between polarization and optical images of the considered BH model.
The rest of this paper is organized as follows. In Sec. {\bf II}, we
briefly define the background of the Schwarzschild BH surrounded by
PFDM configurations as well as the null geodesic equation. In Sec.
{\bf III}, we define the theoretical aspects of relativistic
radiative transfer radiation and the framework of thermal
synchrotron radiation. The mathematical formulation of
the phenomenological model and the corresponding observational
signatures of BH images are investigated in Sec. {\bf IV}. Section
{\bf V} is devoted to analyze the significant results of the ballistic
approximation accretion flow model (BAAF), as well as focusing on
polarization patterns. Finally, we end the paper with a summary.
Throughout this paper, we adopt geometric units with $c = 1 = G$,
where $c$ is the vacuum speed of light and $G$ is the gravitational
constant.

\section{SCHWARZSCHILD Black Hole with Perfect Fluid Dark Matter}
The action of Einstein GR coupled with dark matter, is defined as
follows~\cite{pol24}
\begin{align}
S=\int d^4x\sqrt{-g}\left(\dfrac{\hat{R}}{16 \pi}+
\mathcal{L}_{\text{m}}+\mathcal{L}_{\text{e}}\right),\label{darkmatteraction}
\end{align}
in which $g=det(g_{\mu\nu})$ is the metric determinant, $\hat{R}$
represent the Ricci scalar, $\mathcal{L}_{\text{m}}$ indicates the
dark matter and $\mathcal{L}_{\text{e}}$ is the distribution of dark
energy \cite{pol16}. Varying this action with respect to the metric, the
The Einstein field equation is derived as given below
\begin{align}
\hat{R}_{\mu\nu}-\dfrac{1}{2}g_{\mu\nu}\hat{R}&=8\pi\
T_{\mu\nu}^{ds},\label{DMEinsteinequation}
\end{align}
here $T^{ds}_{\mu\nu}$ represents the fluid distribution of
energy-momentum tensor corresponding to the dark sector, which is the
combination of dark matter and dark energy. The stress tensor for
perfect fluid dark matter is defined as follows \cite{pol16}
\begin{align}
(T^{\mu}~_{\nu})^{m}=\ {\rm
diag}\left(-\rho_{m},0,0,0\right),\label{DMtensor}
\end{align}
which leads to the following form as
\begin{align}
(T^{\mu}~_{\nu})^{ds}=\ {\rm
diag}\left(-\rho,P_r,P_\theta,P_\phi\right),\label{PFDMtensor}
\end{align}
where $\rho=\rho_{m}+\rho_{e}$ illustrates the total energy density
and $P_i$ denotes the total pressure of dark energy. It is worth to
mention here that, the stress tensor $(T^{\mu}~_{\nu})^{ds}$ as
defined in Eq. (\ref{PFDMtensor}) actually contains dark matter and
phantom dark energy, which is not a perfect fluid nor a matter, for a
detail review, the readers can see \cite{supr1,pol24}. Further the
stress tensor $(T^{\mu}~_{\nu})^{ds}$ are defined with constraints
as
\begin{equation}
    \rho=-P_r, \quad P_\theta=P_\phi=\frac{\rho}{2}.
\end{equation}
Static spherically symmetric solutions can be obtained
(\ref{DMEinsteinequation}) as
\begin{align}
ds^2=-f(r)dt^2+\dfrac{1}{f(r)}dr^2+r^2(d\theta^2+\sin^2\theta
d\phi^2)\label{ansatzmetric},
\end{align}
where
\begin{align}
f(r)=1-\dfrac{2M}{r}+\dfrac{\hat{\eta}}{r}\ln\left(\dfrac{r}{|\hat{\eta}|}\right),\label{RNdSPFDMmetric}
\end{align}
in which $M$ is the BH mass and $\hat{\eta}$ represents the PFDM
parameter. From the Einstein equation, we have
\begin{equation}\label{rholabel}
    \rho=-\frac{\hat{\eta}}{8\pi r^3},
\end{equation}
and hence the parameter $\hat{\eta}$ represents the participation of
dark sector to the total energy density, which is the so-called PFDM
parameter. From the perspective of the weak energy condition, it
follows that the energy density $\rho$ should be positive in the
surrounding field, i.e., $\rho \geqslant 0$, which leads to
$\hat{\eta}\leqslant 0$, and hence one can remove the absolute value
symbol $-\hat{\eta}$ from the equation (\ref{RNdSPFDMmetric}) and
rewrite $-\hat{\eta}$ as $\eta$ \cite{supr1}
\begin{align}
f(r)=1-\dfrac{2M}{r}-\dfrac{\eta}{r}\ln\left(\dfrac{r}{\eta}\right),\label{RNdSPFDMmetric}
\end{align}
where $\eta>0$. Now, we are interested in discussing the dynamics of
massless particles around the BH. In this regard, the photon's motion
in the vicinity of a PFDM BH is described by a Lagrangian formalism as
\begin{eqnarray}\label{pn1}
\mathcal{L}=\frac{1}{2}g_{\mu\nu}\dot{x}^{\mu}\dot{x}^{\nu}=
\frac{1}{2}\big(-f(r)\dot{t}^{2}+\frac{\dot{r}^{2}}{f(r)}+r^{2}(\dot{\theta}^{2}+\sin^{2}\theta
\dot{\phi}^{2})\big),
\end{eqnarray}
where $\dot{x}^{\mu}$ is the four-velocity of the photon, and
``dot'' indicates the derivative with respect to the affine
parameter (AP) $\sigma$. For photon $\mathcal{L}=0$ and without loss
of generality, we only consider the motion of photons on the
equatorial plane by setting $\theta=\pi/2$ and $\dot{\theta}=0$.
Since this space-time admits two Killing vector fields,
$\partial_{t}$ and $\partial_{\phi}$, there exists two corresponding
conserved quantities along the trajectory of light, such as the
energy $E$ and the angular momentum $L$, which are defined as below
\begin{eqnarray}\label{pn2}
E=f(r)\frac{dt}{d\sigma}, \quad L=r^{2}\frac{d\phi}{d\sigma}.
\end{eqnarray}
Using Eqs. (\ref{pn1}) and (\ref{pn2}), the four-velocity components
of time, azimuthal angle, and the radial components can be defined
as
\begin{eqnarray}\label{pn3}
\frac{dt}{d\sigma}&=&\frac{1}{b f(r)},
\\\label{pn4}
\frac{d\phi}{d\sigma}&=&\pm\frac{1}{r^{2}},
\\\label{pn5}
\frac{dr}{d\sigma}&=&\sqrt{\frac{1}{b^{2}}-\frac{1}{r^{2}}f(r)},
\end{eqnarray}
in which $b=\frac{L}{E}$, the so-called impact parameter. At the
equatorial plane, one can obtain the expression of the effective
potential in the following form
\begin{eqnarray}
\dot{r}^2=\frac{1}{b^{2}}-V_{eff}(r),\label{p26}
\end{eqnarray}
where
\begin{eqnarray}
V_{eff}(r)=\frac{f(r)}{r^2}. \label{p27}
\end{eqnarray}
At photon sphere, photon motion satisfies $\dot{r}=0$ and
$\ddot{r}=0$. Therefore, the radius of the photon ring satisfies
\begin{equation}
\frac{\partial V_{eff}(r)}{\partial r} = 0. \label{peq:effs}
\end{equation}
To observe the optical image of the BH shadow on the observer's
screen, we consider the zero-angular-momentum observer (ZAMO),
positioned at coordinates ($t_{o},~r_{o},~\theta_{o},~\phi_{o}$),
and a locally orthogonal normalized frame may explain in the
surroundings of the observer, which are
\begin{eqnarray}\nonumber
&&\hat{\eta}_{(t)}=\bigg(\sqrt{-\frac{1}{g_{tt}}},~0,~0,~0\bigg),\;\;\;\;\;\;\;\;\;\;\;
\hat{\eta}_{(r)}=\bigg(-\sqrt{\frac{1}{g_{rr}}},~0,~0,~0\bigg)\\\label{s7}&&
\hat{\eta}_{(\theta)}=\bigg(0,~\sqrt{\frac{1}{g_{\theta\theta}}},~0,~0\bigg),\;\;\;\;\;\;\;\;\;\;\;
\hat{\eta}_{(\phi)}=\bigg(-\sqrt{\frac{1}{g_{\phi\phi}}},~0,~0,~0,\bigg).
\end{eqnarray}

In the framework of ZAMO, the four-momentum of photons can be
defined as $p_{(\alpha)}=p_{\beta}\hat{\eta}^{\beta}_{(\alpha)}$.
From the perspectives of ZAMO, the trajectories of photons can be
described in the observer's frame with the help of celestial
coordinates $(\xi\zeta)$. The tangent vector of the null geodesic
can be described as
\begin{equation}\label{s8}
\dot{S}=|\overrightarrow{OP}|(-\hat{\eta}_{(t)}+\cos\xi\hat{\eta}_{(r)}+\sin\zeta\cos\xi\hat{\eta}_{(\theta)}
+\sin\xi\sin\zeta\hat{\eta}_{(\phi)})
\end{equation}
where $|\overrightarrow{OP}|$ indicates the tangent vector of the
null geodesic at point $O$ in the $3$-dimensional subspace. Closely
followed by \cite{pol25}, the celestial coordinates are defined as

\begin{eqnarray}\label{s9}
\cos\xi=\frac{p^{(r)}}{p^{(t)}},\quad
\tan\zeta=\frac{p^{(\phi)}}{p^{(\theta)}}.
\end{eqnarray}
The relationships between the standard Cartesian coordinate system
($\hat{x},~\hat{y}$) and Celestial coordinates are defined as

\begin{eqnarray}\label{s10}
\hat{x}=-2\tan\frac{\xi}{2}\sin\zeta, \quad
\hat{y}=-2\tan\frac{\xi}{2}\cos\zeta.
\end{eqnarray}
Based on this setup, one can plot the optical images of a BH shadow
and polarization images of a PFDM BH on the observer's screen.

\section{\textbf{Synchrotron Radiation}}

In the extreme plasma environment, the synchrotron radiation is mainly generated by
electrons. Therefore, to accurately define the emission,
absorption, and spin motion of polarized emission in BH space-time,
it is necessary to develop a suitable frame of reference. Closely
followed by~\cite{Broderick:2003fc}, we present the fluid coordinate
system by the four-velocity $u^\mu$, the photon wave four-vector
$k^\mu$, and the local magnetic field $b^\mu$. The corresponding
orthonormal basis vectors are proposed as follows:
\begin{equation}
e_{(t)}^\mu=u^\mu, \quad
e_{(\phi)}^\mu=\frac{k^\mu}{\hat{\omega}}-u^\mu, \quad
e_{(\theta)}^\mu=\frac{1}{\mathcal{S}}\left(b^\mu+\hat{\beta}
u^\mu-g_{1} e_{(\phi)}^\mu\right), \quad
e_{(r)}^\mu=\frac{\epsilon^{\mu \nu \sigma \rho} u_\nu k_\sigma
b_\rho}{\hat{\omega} \mathcal{S}}\,,
\end{equation}
here $\epsilon^{\mu \nu \sigma \rho}$ is the Levi-Civita tensor with
\begin{equation}
b^2=b_\mu b^\mu, \quad \hat{\beta}=u_\mu b^\mu, \quad
\hat{\omega}=-k_\mu u^\mu, \quad g_{1}=\frac{k_\mu
b^\mu}{\hat{\omega}}-\hat{\beta}, \quad
\mathcal{S}=\sqrt{b^2+\hat{\beta}^2-g_{1}^2}\,.
\end{equation}
In this mechanism, all radiations, absorption, and Faraday rotation
coefficients related to the Stokes parameter $U$ will be
disappear. And hence, the remaining Stokes parameters such as $I$,
$Q$, and $V$ associated with non-vanishing emissivities are defined
as follows \cite{code1,code2}
\begin{equation}
    \begin{aligned}
        j_I&=\frac{\sqrt{3}e^3 B \sin{\theta_B}}{4 \pi m_e c^2}\int^\infty_0 \text{d}\gamma N(\gamma)F(\frac{\nu}{\nu_s})\,,\\
        j_Q&=\frac{\sqrt{3}e^3 B \sin{\theta_B}}{4 \pi m_e c^2}\int^\infty_0 \text{d}\gamma N(\gamma)G(\frac{\nu}{\nu_s})\,,\\
        j_V&=\frac{\sqrt{3}e^3 B \sin{\theta_B}}{4 \pi m_e c^2}\int^\infty_0 \text{d}\gamma N(\gamma)\frac{4\cot{\theta_B}}{3\gamma}H(\frac{\nu}{\nu_s})\,,
    \end{aligned}
\end{equation}
where $N(\gamma)$ denotes the electron energy distribution function,
whose form determines the detailed synchrotron emissivity. Here,
$B$ is the magnitude of the local magnetic field,
$\nu$ is the emitted frequency, and $\nu_s = \frac{3 e B
\sin{\theta_B}\, \gamma^2}{4\pi m_e c}$ is the characteristic
synchrotron frequency. The electron Lorentz factor is $\gamma =
1/\sqrt{1-\beta^2}$, where $e$, $m_e$, and $c$ are the elementary
charge, electron mass, and speed of light, respectively. The pitch
angle $\theta_B$ is the angle between the wave vector and the
magnetic field in the fluid rest frame.

The synchrotron functions corresponding to total, linear, and
circularly polarized emission are defined as
\begin{equation}
  \begin{aligned}
      F(x)&=x\int^\infty_x \text{d}y K_{5/3}(y)\,,\\
      G(x)&=xK_{2/3}(x)\,,\\
      H(x)&=\int^\infty_x \text{d}y K_{1/3}(y)+xK_{1/3}(x)\,,
  \end{aligned}
\end{equation}
where $K_n(z)$ denotes the modified Bessel function of second
kind of order $n$.

We adopt a relativistic thermal (Maxwellian) electron distribution,
which is commonly used for astrophysical synchrotron sources:
\begin{equation}
    N(\gamma)=\frac{n_e \gamma^2 \beta}{\theta_e K_2(1/\theta_e)\text{exp}(-\frac{\gamma}{\theta_e})}\,,
\end{equation}
where $n_e$ is the electron number density and $\theta_e = \frac{k_B
T_e}{m_e c^2}$ is the dimensionless electron temperature. Here $k_B$
is the Boltzmann constant and $T_e$ the thermodynamic temperature.
In the ultra-relativistic regime ($\beta \approx 1$, $\theta_e \gg
1$), the modified Bessel function can be approximated by
$K_2(1/\theta_e) \simeq 2\theta_e^2$. Introducing $z \equiv \gamma /
\theta_e$, the emissivities become
\begin{equation}
    \begin{aligned}
        j_I&=\frac{\sqrt{3} n_e e^3 B \sin{\theta_B}}{8 \pi m_e c^2}\int^\infty_0 \text{d}z z^2\text{exp}(-z)F(\frac{\nu}{\nu_s})\,,\\
        j_Q&=\frac{\sqrt{3} n_e e^3 B \sin{\theta_B}}{8 \pi m_e c^2}\int^\infty_0 \text{d}z z^2\text{exp}(-z)G(\frac{\nu}{\nu_s})\,,\\
        j_V&=\frac{\sqrt{3} n_e e^3 B \sin{\theta_B}\cot{\theta_B}}{6 \pi m_e \theta_e c^2}\int^\infty_0 \text{d}z z\text{exp}(-z)H(\frac{\nu}{\nu_s})\,.
    \end{aligned}
\end{equation}

Defining $x=(\nu/\nu_s) z^2$, the emissivities can be written compactly as
\begin{equation}
    \begin{aligned}
        j_I&=\frac{n_e e^2\nu}{2\sqrt{3}c\theta_e^2}I_I(x)\,,\\
        j_Q&=\frac{n_e e^2\nu}{2\sqrt{3}c\theta_e^2}I_Q(x)\,,\\
        j_V&=\frac{2n_e e^2\nu\cot{\theta_B}}{3\sqrt{3}c\theta_e^3}I_V(x)\,,
    \end{aligned}
\end{equation}
where $x \equiv \nu / \nu_c$ is the ratio of the emitted photon
frequency to the characteristic frequency of the system
$\nu_c=\frac{3 e B \sin{\theta_B} \theta_e^2}{4\pi m_e c}$. The
dimensionless thermal synchrotron integrals are
\begin{equation}
\begin{aligned}
     I_I(x)&=\frac{1}{x}\int^\infty_0 \text{d} z z^2 \text{exp}(-z) F\left(\frac{x}{z^2}\right)\,,\\
     I_Q(x)&=\frac{1}{x}\int^\infty_0 \text{d} z z^2 \text{exp}(-z) G\left(\frac{x}{z^2}\right)\,,\\
     I_V(x)&=\frac{1}{x}\int^\infty_0 \text{d} z z \text{exp}(-z) H\left(\frac{x}{z^2}\right)\,.
\end{aligned}
\end{equation}
For a hot electron plasma, the absorption coefficients obey
Kirchhoff's law, such as
\begin{equation}
a_\nu=\frac{j_\nu}{B_\nu}\,,
\end{equation}
in which $B_\nu$ denotes the Planck black body radiation function.
And then, the final expression of Faraday rotation coefficients is
defined as
\begin{equation}
\begin{aligned}
& r_Q=\frac{n_e e^4 B^2 \sin ^2 \theta_B}{4 \pi^2 m_e^3 c^3 \nu^3} f_m(X)+\left(\frac{K_1\left(\theta_e^{-1}\right)}{K_2\left(\theta_e^{-1}\right)}+6 \theta_e\right), \\
& r_V=\frac{n_e e^3 B \cos \theta_B}{\pi m_e^2 c^2 \nu^2}
\frac{K_0\left(\theta_e^{-1}\right)-\Delta
J_5(X)}{K_2\left(\theta_e^{-1}\right)}\,,
\end{aligned}
\end{equation}
with
\begin{equation}
X=\frac{1}{\left(\frac{3}{2 \sqrt{2}} \times 10^{-3}
\frac{\nu}{\nu_c}\right)^{1 / 2}}\,.
\end{equation}
Hence, the emissivity, absorption, and Faraday coefficients
primarily depend on the electron number density, magnetic field
strength, field-wave vector angle, and electron temperature. In
order to observe the BH image at the observer's position, we vibrate
the radiation from the light source to the observer's frame. To
explain the relation between light rays and matter in radiative
transfer, we impose the tensor form of the covariant radiative
transfer equation, which has the following expression as
\cite{Gammie}
\begin{equation}
\label{rteq} k^\mu \nabla_\mu \bar{\mathcal{S}}^{\alpha
\beta}=\mathcal{J}^{\alpha \beta}+H^{\alpha \beta \mu \nu}
\bar{\mathcal{S}}_{\mu \nu}\,,
\end{equation}
here $\bar{\mathcal{S}}^{\alpha \beta}$ corresponds to the
polarization tensor, $k^\mu$ is the photon wave vector,
$\mathcal{J}^{\alpha \beta}$ characterizes emission from the source,
and $H^{\alpha \beta \mu \nu}$ encodes absorption and Faraday
rotation effects. The corresponding numerical solution of Eq.
(\ref{rteq}) can be derived from the public code
\textsc{Coport}~1.0~\cite{code1}. Based on \textsc{Coport}~1.0, we
exploit the gauge invariance of $\bar{\mathcal{S}}^{\alpha\beta}$ to
simplify computations in a properly chosen parallel-transported
frame. In this regard, Eq.~\eqref{rteq} can be classified into two
portions. The first part corresponds to the gravitational effects,
such as
\begin{equation}
    k^\mu \Delta_\mu f^a=0\,,\,\,f^ak_a=0\,,
\end{equation}
where $f^\mu$ is a normalized spacelike vector orthogonal to
$k^\mu$. The impact of plasma accretion can be investigated through
the second part, which is

\begin{equation}
\frac{\text{d}\hat{S}}{\text{d} \sigma}=R(\chi) J-R(\chi)
\mathcal{M} R(-\chi) \hat{S}\,,
\end{equation}

where
\begin{equation}
\hat{S}=\left(\begin{array}{l}
\mathcal{I} \\
Q \\
\mathcal{U} \\
\mathcal{V}
\end{array}\right), \quad J=\frac{1}{\nu^2}\left(\begin{array}{l}
j_I \\
j_Q \\
j_U \\
j_V
\end{array}\right), \quad \mathcal{M}=\nu\left(\begin{array}{cccc}
a_I & a_Q & a_U & a_V \\
a_Q & a_I & r_V & -r_U \\
a_U & -r_V & a_I & r_Q \\
a_V & r_U & -r_Q & a_I
\end{array}\right) \,,
\end{equation}
in which
$\mathcal{I}=\mathcal{Q}=I(\nu)^{-3}=\mathcal{U}=\mathcal{V}$. The
matrix $R(\chi)$ indicates rotation between the synchrotron
emission frame and the parallel-transported reference frame as
\begin{equation}
\label{rotationmatrix} R(\chi)=\left(\begin{array}{cccc}
1 & & & \\
& \cos (2 \chi) & -\sin (2 \chi) & \\
& \sin (2 \chi) & \cos (2 \chi) & \\
& & & 1
\end{array}\right)\,,
\end{equation}
where $\chi$ is the rotation angle, illustrating the angle between
the reference vector $f^\mu$ and the magnetic field $B^\mu$:
\begin{equation}
\chi=\operatorname{sign}\left(\epsilon_{\mu \nu \alpha \beta} u^\mu
f^\nu B^\rho k^\sigma\right) \arccos \left(\frac{\bar{P}^{\mu \nu}
f_\mu B_\nu}{\sqrt{\left(\bar{P}^{\mu \nu} f_\mu
f_\nu\right)\left(\bar{P}^{\alpha \beta} B_\alpha
B_\beta\right)}}\right)\,,
\end{equation}
and $\bar{P}^{\mu\nu}=g^{\mu\nu}-\frac{k^\mu
k^\nu}{\hat{\omega}^2}+\frac{u^\mu k^\nu+k^\mu u^\nu}{\hat{\omega}}$
is the induced metric on the transverse subspace. At the observers
position, the Stokes parameters are projected onto the screen with
the help of the same rotation matrix, as defined in
Eq.~\eqref{rotationmatrix}, having the corresponding rotation angle
\begin{equation}
\chi_o=\operatorname{sign}\left(\epsilon_{\mu \nu \rho \sigma}
u_o^\mu f^\nu d^\rho k^\sigma\right) \arccos
\left(\frac{\bar{P}^{\mu \nu} f_\mu d_\nu}{\sqrt{\left(\bar{P}^{\mu
\nu} f_\mu f_\nu\right)\left(\bar{P}^{\alpha \beta} d_\alpha
d_\beta\right)}}\right)\,,
\end{equation}
where $u_o^\mu$ is the observer's four-velocity and $d^\mu$ denotes
the $y$-axis direction of the observer's screen. In this work, we
consider $d^\mu=-\partial_\theta^\mu$. And then, the corresponding
projected Stokes parameters have the following expressions as
\begin{equation}
\mathcal{I}_o=\mathcal{I}, \quad \mathcal{Q}_o=\mathcal{Q} \cos
2\chi_o-\mathcal{U} \sin 2\chi_o, \quad \mathcal{U}_o=\mathcal{Q}
\sin 2\chi_o+\mathcal{U} \cos 2\chi_o, \quad
\mathcal{V}_o=\mathcal{V}\,.
\end{equation}
The observed Stokes parameters interpret the polarization state of
the radiation. The total intensity is given by $\mathcal{I}_o$,
while $\mathcal{V}_o$ denotes circular polarization with $+$ and $-$
values, which correspond to left and right-handed circular
polarization, respectively. The linear polarization intensity and
electric-vector position angle (EVPA) is given by
\begin{equation}
P_o=\sqrt{\mathcal{Q}_o^2+\mathcal{U}_o^2}, \quad
\Phi_{\mathrm{EVPA}}=\frac{1}{2} \arctan
\frac{\mathcal{U}_o}{\mathcal{Q}_o}\,.
\end{equation}
Next, we are going to discuss the visual properties of a BH in the
vicinity of PFDM with the help of two realistic models of
geometrically thick accretion flows, such as a phenomenological RIAF
model and an analytic thick-disk model, which is introduced by Hou
et al. \cite{hou2024new}. For convenience, we shall refer to the Hou
disk model as the ballistic approximation accretion flow model
(BAAF).

\section{Phenomenological Model}

Now we consider cylindrical coordinates, such as
$\bar{R}=r\sin\theta$ represents the cylindrical radius and
$z=r\cos\theta$, calculate the vertical distance from the equatorial
plane $\theta=\pi/2$. Closely followed by
\cite{broderick2011evidence}, we express the mathematical expression
of the analytically radiatively inefficient accretion flow (RIAF) disk
model, the radial and vertical profiles of the number density and
the temperature configurations have the following form
\begin{equation}
    \bar{n}_e = n_h\left(\frac r{r_h}\right)^2\exp\left(-\frac{s^2}{2(\Gamma \bar{R})^2}\right),\quad T_e = T_h\left(\frac r{r_h}\right),
\end{equation}
here $r_h$ interpret the BH outer horizon, $n_h$ and $T_h$
represents the corresponding electron number density and temperature
at the horizon, respectively. Moreover, the parameter $\Gamma$ used
to calculate the disk thickness. The cold magnetization parameter
$\hat{\lambda}$ measures the strength of the local magnetic field, which
is defined as
\begin{equation}
B=\sqrt{\hat{\lambda}\hat{\rho}},
\end{equation}
where $\hat{\rho}=\bar{n}_e(m_{p}c^{2})$ is the dimensionless fluid
mass density. Generally, the parameter $\hat{\lambda}$ can be
expressed as a spatially varying distribution such as,
$\hat{\lambda}=\hat{\lambda}_{h}r_{h}/r$ \cite{pu2016effects}, where
$\hat{\lambda}_{h}$ is its value at the event horizon and for
conveniently, we fix $\hat{\lambda}=0.1$ in all cases. Since, the
motion of fluid in the present mechanism is relatively free from any
constraints and hence, we assume three representative kinematic
configurations, which are defined as given below

\textbf{(i)~Orbiting motion} For circular motion around the BH, the
fluid four-velocity has two components, such as $u^{t}$ and
$u^{\phi}$, which are defined as given below
\cite{gold2020verification}
\begin{equation}
    u^\mu=u^t \{ 1, 0, 0, \varpi \},\label{eq:om}
\end{equation}
with
\begin{equation}
    u^t = \sqrt{-\frac{1}{g_{tt} + g_{\phi\phi} \varpi^2}}
    ,\quad\varpi = -\frac{g_{tt} \ell}{g_{\phi\phi}},\quad \ell =-\frac{u_\phi}{u_t}= \ell_0\frac{\bar{R}^{3/2}}{g_{1}+\bar{R}}.
\end{equation}
where $\ell$ is the angular momentum density, and $\ell_{0}$ and
$g_{1}$ are model free parameters. In the present work, we fix
$\ell_0=g_{1}=1$ \cite{gold2020verification}.

\textbf{(ii)~Infalling motion}

Assuming that the fluid is at rest at infinity, such as $u_{t}=-1$,
the four-velocity is given by
\begin{equation}
    u^\mu=\{-g^{tt},-\sqrt{-(1+g^{tt})g^{rr}},0,0\},\label{eq:im}
\end{equation}
which corresponds to a special case of the conical solution.

\textbf{(iii)~Combined motion}

The combination of both orbiting and infalling motion, so-called the
combined motion \cite{pu2016effects}, and we have
\begin{equation}
u^\mu = (u^t,\, u^r,\, 0,\, u^\phi)\,,
\end{equation}
where
\begin{equation}
\varpi = \frac{u^\phi}{u^t} = \varpi_o + \beta_1(\varpi_i -
\varpi_o), \qquad u^r = u_o^r + \beta_2 (u_i^r - u_o^r)\,,
\end{equation}
and
\begin{equation}
\varpi_o = -\frac{g_{tt} \ell}{g_{\phi\phi}}, \qquad \varpi_i = 0,
\qquad u_o^r = 0, \qquad u_i^r = \sqrt{-(1 + g^{tt}) g^{rr}}\,,
\end{equation}
where the subscripts~$i$~and~$o$ corresponds to the infalling and
orbiting components, respectively. The range of parameters $\beta_1$
and $\beta_2$ are lies in $(0,1)$, which is used to controlling the
relative weights of the two components: smaller values correspond to
predominantly orbiting motion, while larger values indicate a more
infalling flow. Where we fix $\beta_1=\beta_2=0.2$ throughout this
manuscript. The temporal component $u^t$ is then calculated by
$u^2=-1$.
\begin{equation}
u^t=\sqrt{-\frac{1+g_{r r}\left(u^r\right)^2}{g_{t t}+\varpi^2
g_{\phi \phi}}}\,.
\end{equation}
Next, let us consider the magnetic field. Under the assumption of
ideal magnetohydrodynamic (MHD),
\begin{equation}
    u_{\mu}\mathcal{F}^{\mu\nu}=0\,.
\end{equation}
the electric field in the static frame of the fluid, $e^\nu =0=u_\mu
\mathcal{F}^{\mu\nu}$, and the magnetic field $b^\nu = u_\mu
{^*\mathcal{F}}^{\mu\nu}$ is orthogonal to the fluid four-velocity,
$b_\mu u^\mu = 0$. For all aforementioned fluid scenarios, we assume
a purely toroidal magnetic field,
\begin{equation}
\label{toroidalmag}
    b^\mu\sim(\ell,0,0,1)\,.
\end{equation}
Depending on whether the emissivity depends on the angle between the
magnetic field and the radiated photons, we differentiate between the two
classes of fluids, such as isotropic and anisotropic emission. In
the isotropic case, the emissivity is independent of this angle, and
only the magnetic field strength needs to be specified. Hence, we
impose an angle-averaged emissivity, which is expressed as
\begin{equation}
\bar{j}_{\nu}=\frac{1}{2}\int_{0}^{\pi}j_{\nu}\sin\theta_{B}d\theta_{B},
\end{equation}
with the corresponding fitting formula
\cite{leung2011numerical,mahadevan1996harmony}
\begin{eqnarray}
\bar{j}_{\nu}=\frac{\bar{n}_{e}e^{2}\nu}{2\sqrt{3}c\theta^{2}_{e}}I(x),
\quad x=\frac{\nu}{\nu_{c}}, \quad
\nu_{c}=\frac{3eB\theta^{2}_{e}}{4\pi m_{e}c},
\end{eqnarray}
where $I(x)$ is approximated defined as given below
\cite{leung2011numerical,mahadevan1996harmony}
\begin{equation}
    I(x)=\frac{4.0505}{x^{1/6}}\left(1+\frac{0.4}{x^{1/4}}+\frac{0.5316}{x^{1/2}}\right)\exp\left(-1.8899x^{1/3}\right).
\end{equation}
For the anisotropic framework, the direction of the magnetic field
must be accounted for. Adopting the toroidal field configuration in
Eq.~\eqref{toroidalmag}, and the corresponding fitting function is
defined as \cite{mahadevan1996harmony}
\begin{equation}
I(x)=2.5651\left(1+1.92 x^{-1 / 3}+0.9977 x^{-2 / 3}\right) \exp
\left(-1.8899 x^{1 / 3}\right)\,.
\end{equation}
Based on the aforementioned mechanism, next, we are going to
illustrate the BH images plotted from the phenomenological RIAF
model.
\subsection{\textbf{Isotropic Radiation Case}}
It is widely acknowledged that the thermal synchrotron radiation is
intrinsically anisotropic, with its emissivity strongly dependent on
the emission direction. This directional dependence is evaluated by
the pitch angle $\theta_{B}$, which is defined as the angle between
the photon wave vector and the magnetic field in the rest frame of
the fluid. To quantify the influence of this anisotropy on BH
images, it is necessary to compare the results with those obtained
from an isotropic emission profile, which serves as a baseline for
assessing anisotropy induced effects.

\begin{figure}[H]\centering
\subfigure[$\eta=0.1,\theta_o=0^\circ$]{\includegraphics[scale=0.4]{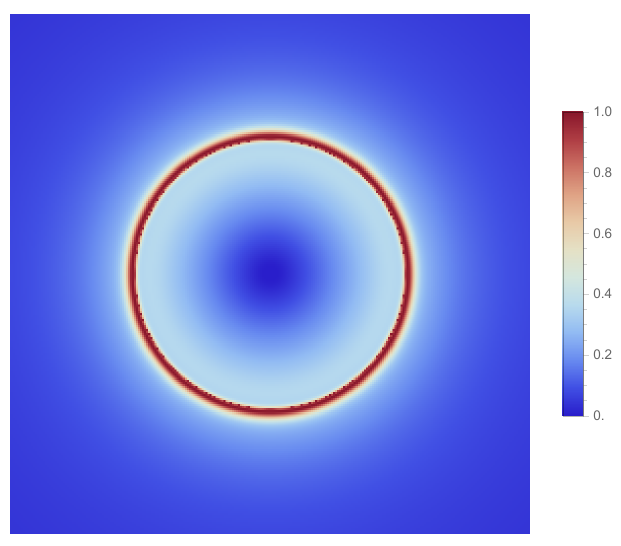}}
\subfigure[$\eta=0.1,\theta_o=17^\circ$]{\includegraphics[scale=0.4]{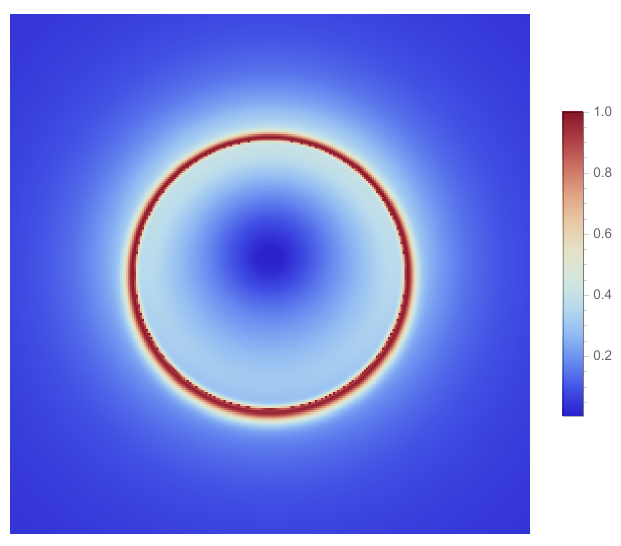}}
\subfigure[$\eta=0.1,\theta_o=70^\circ$]{\includegraphics[scale=0.4]{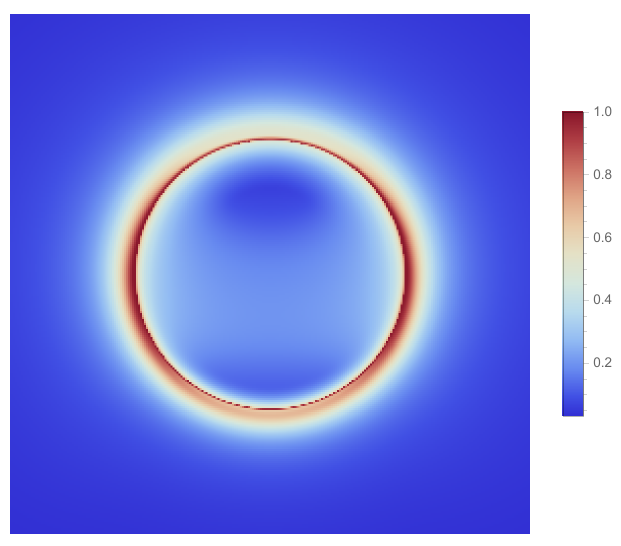}}
\subfigure[$\eta=0.3,\theta_o=0^\circ$]{\includegraphics[scale=0.4]{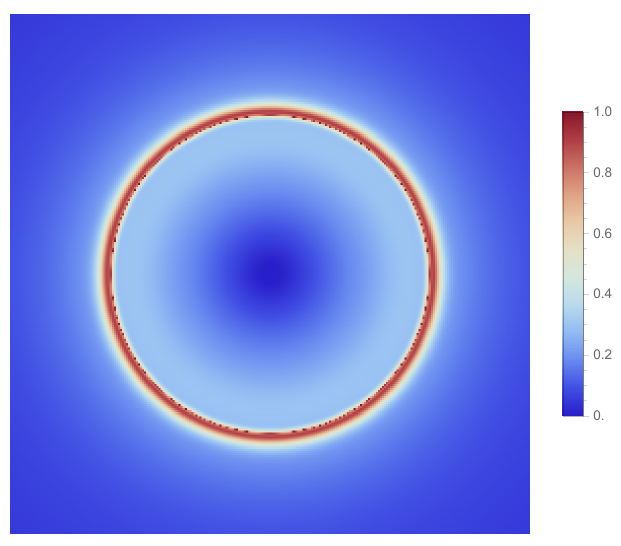}}
\subfigure[$\eta=0.3,\theta_o=17^\circ$]{\includegraphics[scale=0.4]{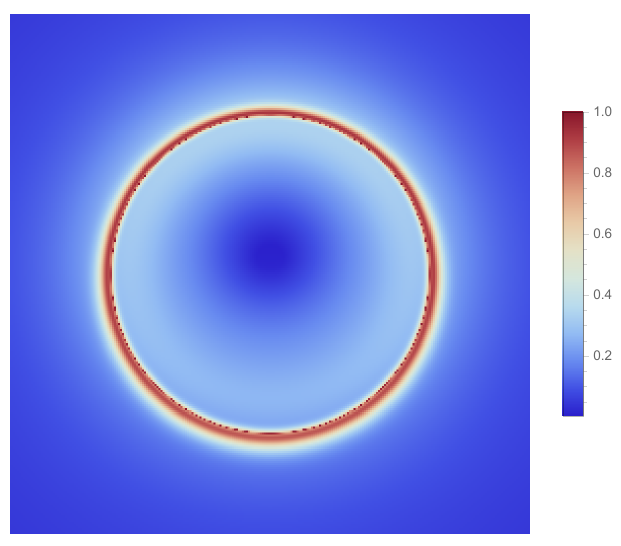}}
\subfigure[$\eta=0.3,\theta_o=70^\circ$]{\includegraphics[scale=0.4]{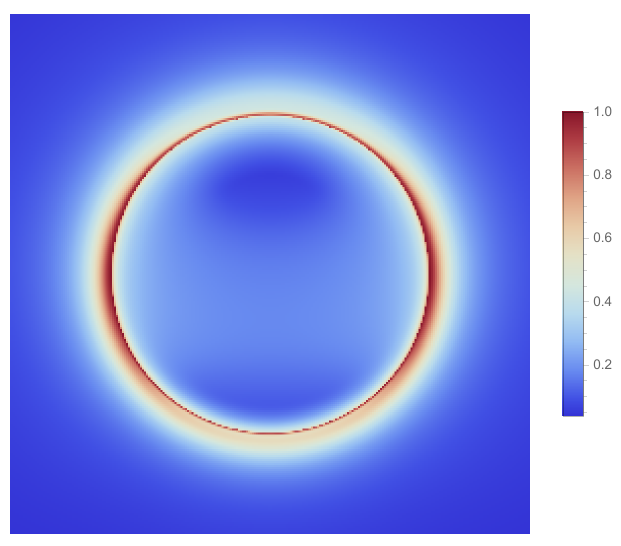}}
\subfigure[$\eta=0.5,\theta_o=0^\circ$]{\includegraphics[scale=0.4]{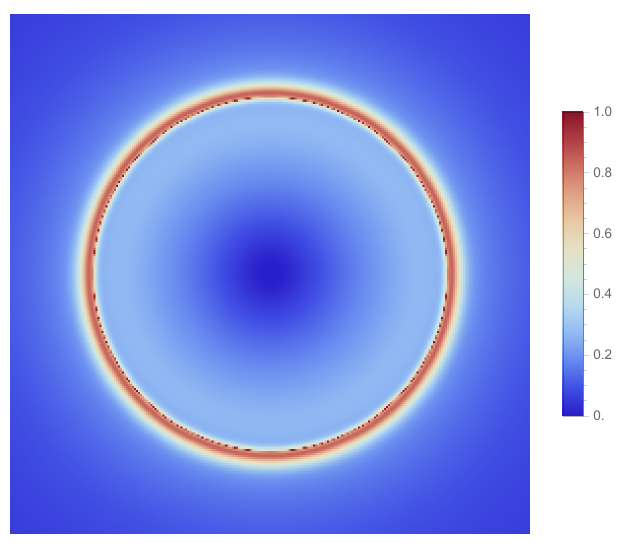}}
\subfigure[$\eta=0.5,\theta_o=17^\circ$]{\includegraphics[scale=0.4]{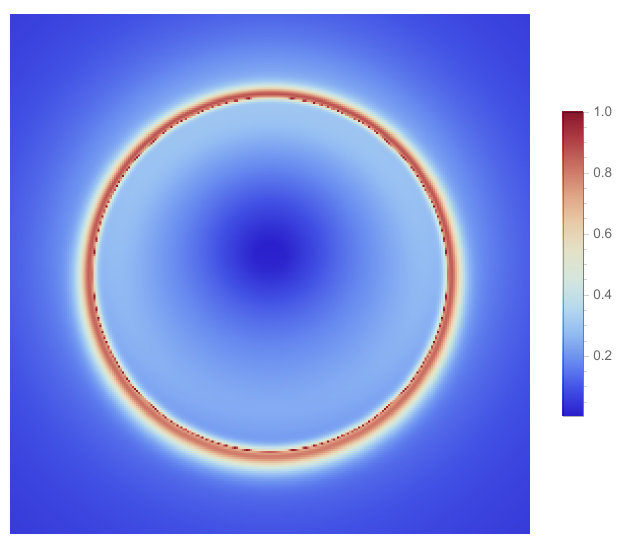}}
\subfigure[$\eta=0.5,\theta_o=70^\circ$]{\includegraphics[scale=0.4]{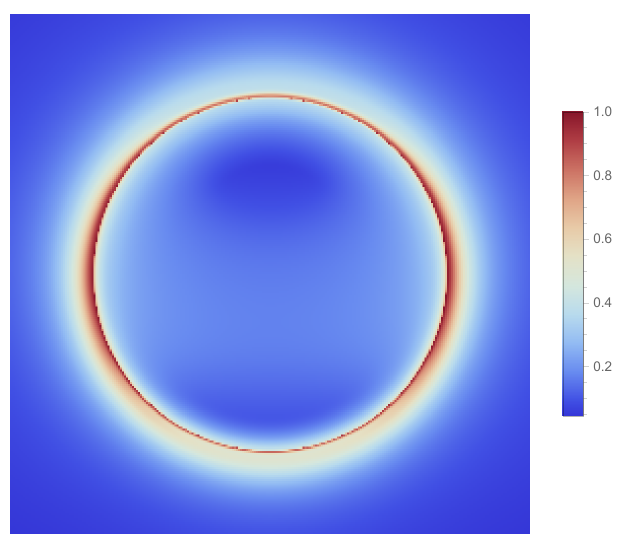}}
\caption{Effects of $\eta$ and $\theta_o$ on the RIAF model with
isotropic radiation at the observation frequency of
$230\,\mathrm{GHz}$, with the accretion flow in the infalling
motion.}\label{fig1}
\end{figure}

\begin{figure}[H]
       \centering
       \subfigure[Horizontal, isotropic]{\includegraphics[scale=0.8]{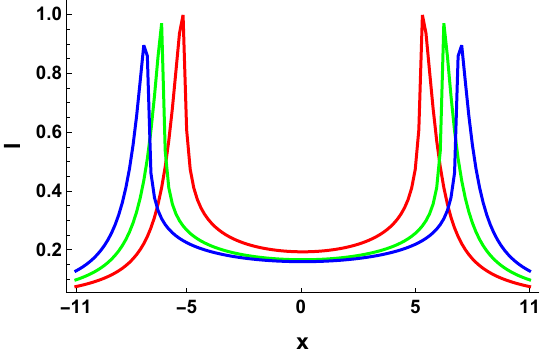}}
       \subfigure[Vertical, isotropic]{\includegraphics[scale=0.8]{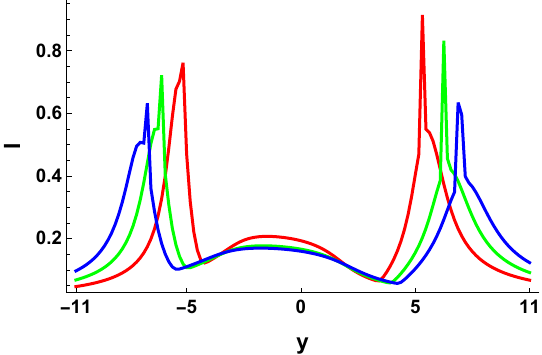}}
\caption{The intensity cuts with $\theta_o=70^\circ$ and the
accretion flow in the infalling motion. The red, green, and blue
curves correspond to $\eta=0.1,\,0.3,\,0.5$,
respectively.}\label{fig2}
\end{figure}

In Fig. \textbf{\ref{fig1}}, we illustrate the intensity maps of
the Schwarzschild BH in the presence of PFDM configurations with
RIAF model under isotropic radiation. The accretion flow follows an
infalling motion, and the observing frequency corresponds to
$230$GHz. From left to right, each panel corresponds to different
values of inclination angle $\theta_{o}$, such as
$\theta_{o}=0^\circ,~17^\circ,~70^\circ$, while from top to bottom,
the PFDM parameter takes the values $\eta=0.1,~0.3,~0.5$,
respectively. For a comparative analysis, we also depicted the
intensity distribution in Fig. \textbf{\ref{fig2}} along the
observer's screen $x$-and $y$-intercepts, where the red, green, and
blue curves correspond to $\eta=0.1,\,0.3,\,0.5$, respectively. All
these images show that a brightest photon ring, which corresponds
to the peak curves of Fig. \textbf{\ref{fig2}}. These features come
from higher-order images, so-called photons that move around the BH
one or more times before approaching the observer's screen,
constituting a direct demonstration of gravitational lensing. The
region outside the ring illustrates the primary image, formed by
photons reaching the observer directly from the accretion disk. The
dark interior regions inside the images indicate the BH event
horizon. For geometrically thin disks, the BH horizon generates a
pronounced inner shadow, which may be observable with the EHT
\cite{chael1}. On the other hand, for geometrically thick disks,
emission from the off-equatorial zones can obscure the horizon
boundary, thereby making the inner shadow barely visible.
Comparing the columns of Fig. \textbf{\ref{fig1}} and Fig.
\textbf{\ref{fig2}}, we notice that for a fixed inclination angle
$\theta_{o}$ and increasing $\eta$ from top to bottom, results in
increasing both the bright ring and the central dark region without
altering their shapes, since the parameter $\eta$ significantly
changes the values of the BH horizon as well as the corresponding critical
impact parameter $b_{c}$. On the other hand, for a fixed value of
$\eta$, and changes in $\theta_{o}$ from left to right, we observe
that there appears a bright ring and the central dark region,
constantly lies in the centre and shows an isotropic nature. When
$\theta_{o}=17^\circ$, a clear up and down asymmetry in the bright
ring emerges, and when $\theta_{o}=70^\circ$, two distinct dark
regions exhibit inside the bright ring, with the upper one more
darker compared to the bottom. The insistent left to right brightness
symmetry originates from the spherical symmetry of the spacetime and
the fluid motion is infalling accretion. On the contrary, the
up-down brightness asymmetry comes from the equatorial symmetry of
the thick disk, such as for observers near the equatorial plane,
high-latitude emission partially fills in the inner shadow, while
near the poles, insufficient photons can reach the
observer's screen. Further, Fig. \textbf{\ref{fig2}} depicts these
effects through horizontal and vertical intensity cuts, where the
variation in $\eta$ significantly changes the peak and the widths
between the curves.

\subsection{\textbf{Anisotropic Radiation Case}}
\begin{figure}[H]\centering
    \subfigure[$\eta=0.1,\theta_o=0^\circ$]{\includegraphics[scale=0.4]{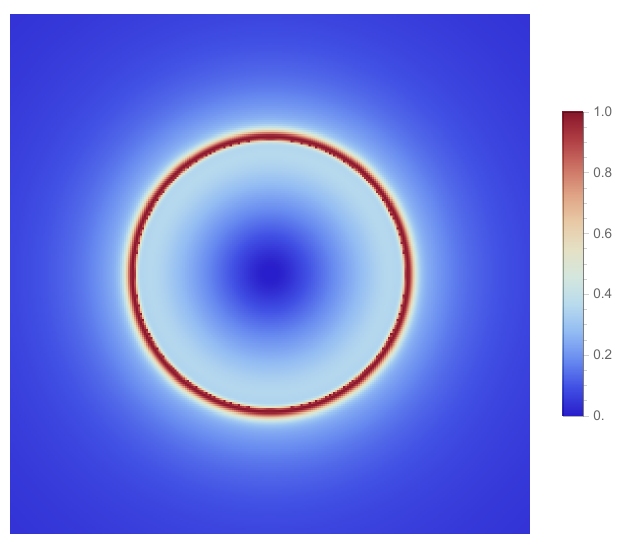}}
    \subfigure[$\eta=0.1,\theta_o=17^\circ$]{\includegraphics[scale=0.4]{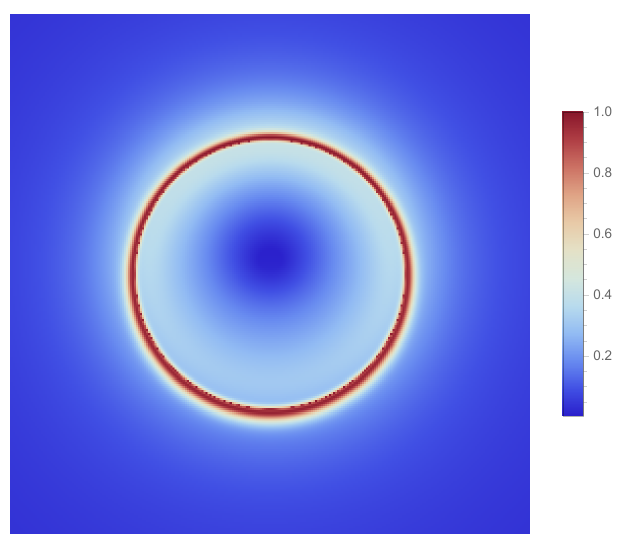}}
    \subfigure[$\eta=0.1,\theta_o=70^\circ$]{\includegraphics[scale=0.4]{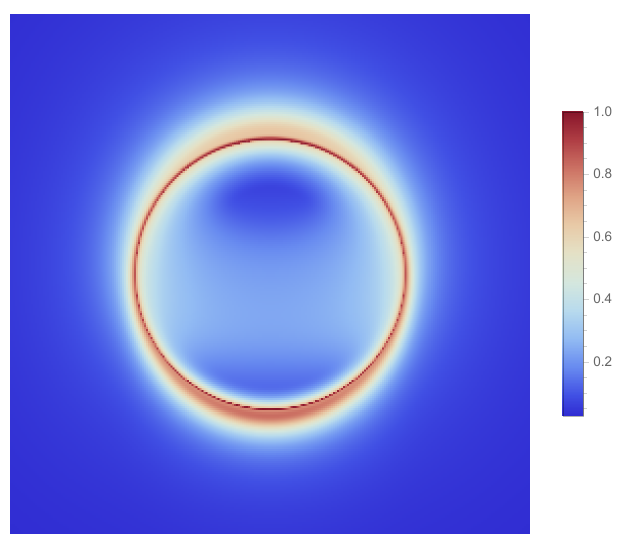}}

    \subfigure[$\eta=0.3,\theta_o=0^\circ$]{\includegraphics[scale=0.4]{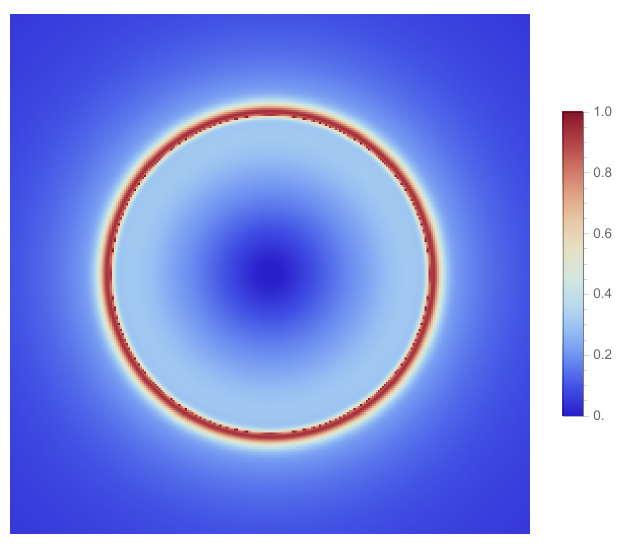}}
    \subfigure[$\eta=0.3,\theta_o=17^\circ$]{\includegraphics[scale=0.4]{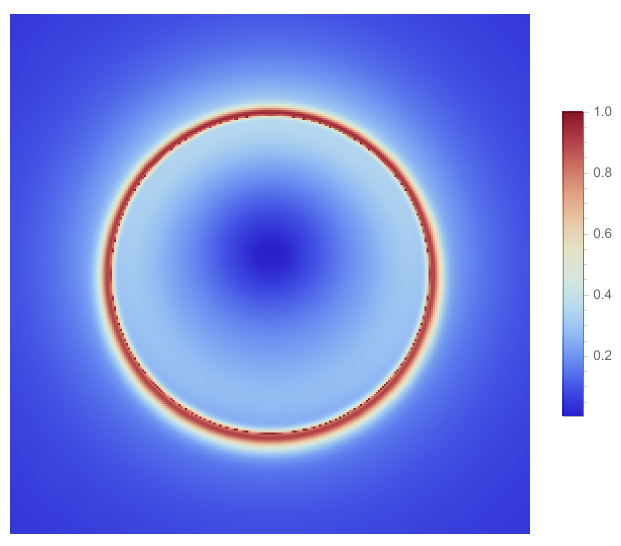}}
    \subfigure[$\eta=0.3,\theta_o=70^\circ$]{\includegraphics[scale=0.4]{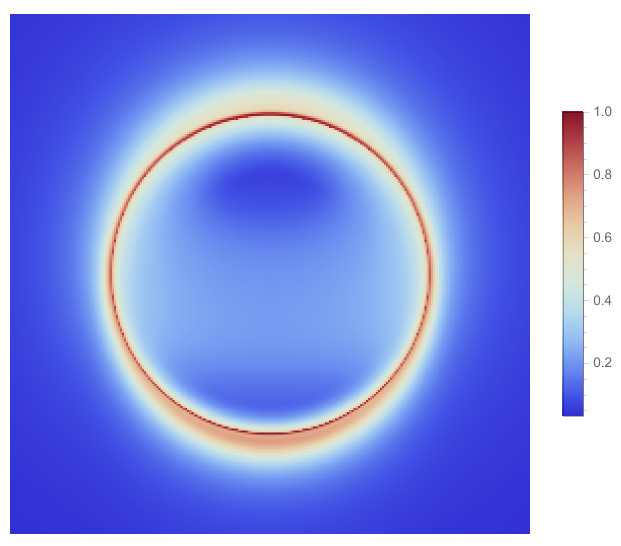}}

    \subfigure[$\eta=0.5,\theta_o=0^\circ$]{\includegraphics[scale=0.4]{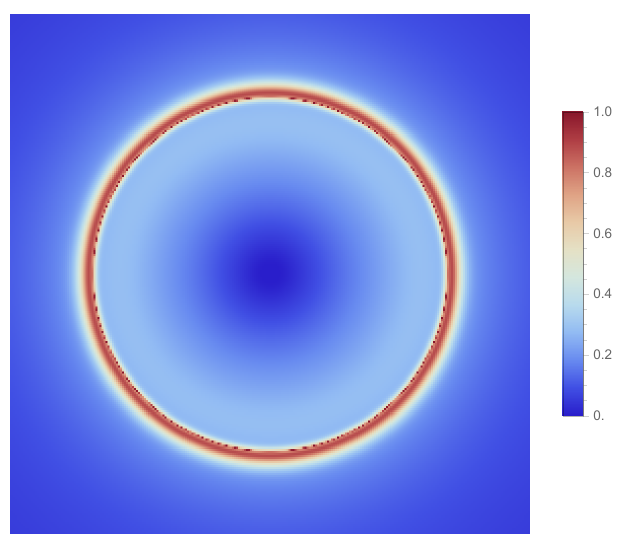}}
    \subfigure[$\eta=0.5,\theta_o=17^\circ$]{\includegraphics[scale=0.4]{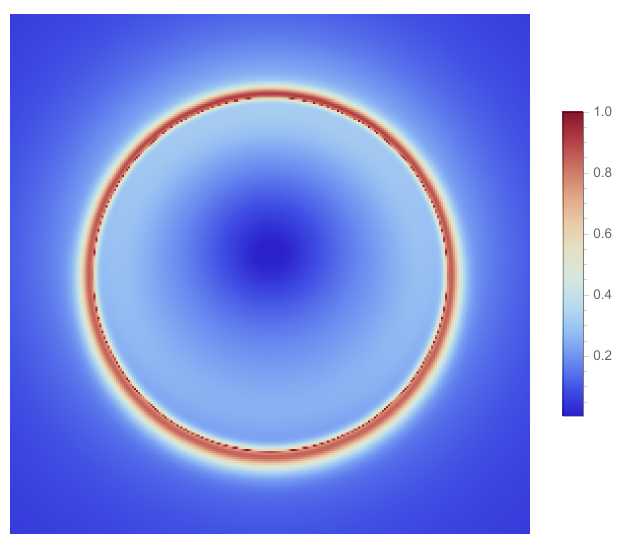}}
    \subfigure[$\eta=0.5,\theta_o=70^\circ$]{\includegraphics[scale=0.4]{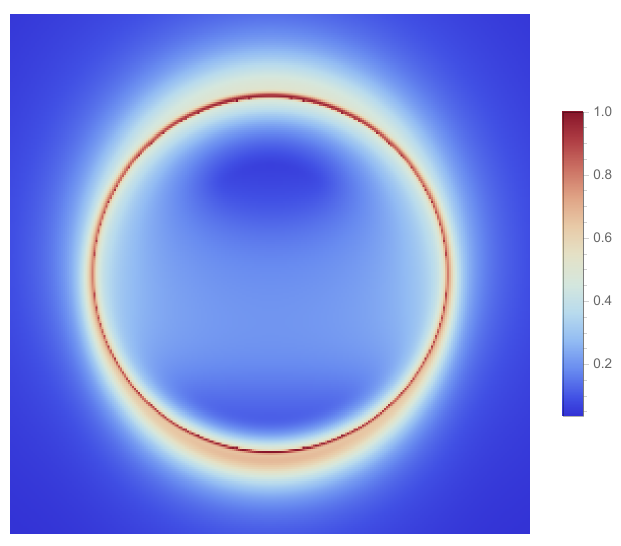}}
\caption{Effects of $\eta$ and $\theta_o$ on the RIAF model with
anisotropic radiation at $230\,\mathrm{GHz}$, with the accretion
flow in the infalling motion.}\label{fig3}
\end{figure}

\begin{figure}[H]
\centering \subfigure[Horizontal,
anisotropic]{\includegraphics[scale=0.8]{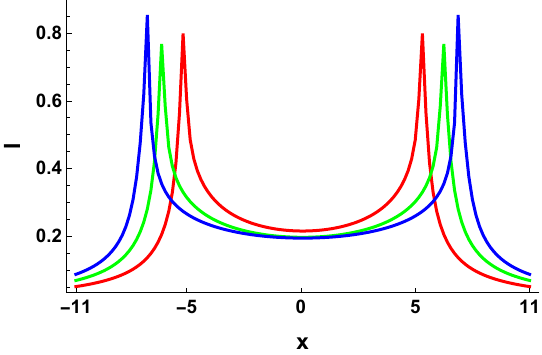}}
\subfigure[Vertical,
anisotropic]{\includegraphics[scale=0.8]{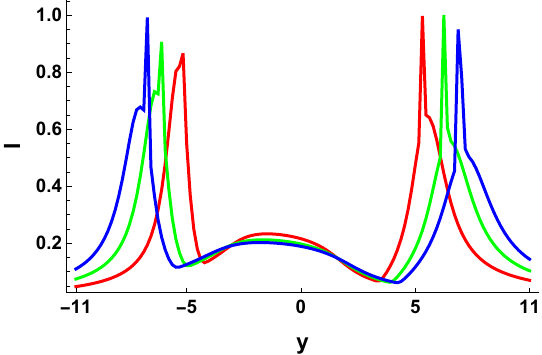}} \caption{The
intensity cuts with $\theta_o=70^\circ$ and the accretion flow in
the infalling motion. The red, green, and blue curves correspond to
$\eta=0.1,\,0.3,\,0.5$, respectively.}\label{fig4}
\end{figure}

Now, we investigate the optical features of an anisotropic synchrotron
emission, considering a toroidal magnetic field distribution as
given in Eq.~\eqref{toroidalmag}. In Fig. \textbf{\ref{fig3}}, we
exhibit the observed intensity of the considered BH model under
RIAF framework at observed frequency $230$GHz with infalling
accretion flow, where the numerical values of the parameters are the
same, as those defined in Fig. \textbf{\ref{fig1}}. For quantitative
comparison, the horizontal and vertical intensity cuts are presented
in Fig. \textbf{\ref{fig4}}. The overall morphology closely aligns
with the isotropic scenario, as observed in Fig.
\textbf{\ref{fig1}}, presenting a pronounced bright ring encircling
a central region, both of which expand with the aid of $\eta$. The
increasing values of $\theta_{o}$, produced the nonuniform
brightness distribution, and two dark regions emerge within the
ring. A notable distinction of the anisotropic scenario is the
development of a vertically elongated and elliptical ring structure
at higher inclinations. This asymmetry comes from the angular
dependence of synchrotron emissivity, for photons radiated from the
upper and lower zones of the disk propagate closely perpendicular to
the magnetic field, increasing the emission and
stretching the ring vertically.

Further, the intensity cuts diagrams interpret the two prominent
peaks corresponding to the higher-order images, whereas the regions
outside the peaks correspond to the primary image. The increasing
values of $\eta$ enhanced the gap between the curves, which
corresponds to increasing the size of the higher-order images. For
the intensity distribution along the $x-$axis, no region with zero
intensity is observed, which is due to the radiation from beyond the
equatorial plane and the observer inclination angle
$\theta_o=70^\circ$. And the intensity distribution along the
$y-$axis, two local minima are exhibited between the two peaks,
corresponding to the event horizon. A local maximum exhibits between
the two minima, which is also due to the influence of radiation from
beyond the equatorial plane.

\section{BAAF Disk Model}
Now, we turn to a more analytically tractable model, the ``BAAF''
model, which is a more realistic model, to further explore the
geometric and radiative properties of a thick accretion disk. The Hou
disk model~\cite{hou2024new} successfully describes a steady,
axisymmetric accretion configuration in which the flow satisfies
$u^\theta \equiv 0$, indicating that the streamlines are best-fit to
constant $\theta$ surfaces. Under this mechanism, the mass
conservation equation reduces to
\begin{equation}
\frac{\mathrm{d}}{\mathrm{~d} r}\left(\sqrt{-g} \bar{\rho}
u^r\right)=0, \quad \Longrightarrow \bar{\rho}=\bar{\rho}_0
\frac{\left.\sqrt{-g} u^r\right|_{r=r_0}}{\sqrt{-g} u^r}\,,
\end{equation}
where $\bar{\rho}_0=\bar{\rho}(r_0)$ represents the static mass
density at a reference radius $r_0$, typically chosen to be the
event horizon, $r_0=r_{h}$. Moreover, the projection of the
energy-momentum conservation equation $\nabla_\mu T^{\mu\nu}=0$
along the four-velocity yields
\begin{equation}
\label{energy} \mathrm{d}\Pi=\frac{\Pi+p}{\bar{\rho}} \mathrm{d}
\bar{\rho}\,,
\end{equation}
in which $\Pi$ represents the internal energy density of the fluid
configuration. Explaining the proton-to-electron temperature ratio
$z=T_{p}/T_{e}$, which quantifies the relative thermal states of the
two species in the plasma, the internal energy density satisfies
\begin{equation}\label{xi}
\Pi=\bar{\rho}+\bar{\rho} \frac{3}{2}(z+2) \frac{m_e}{m_p}
\theta_e\,,
\end{equation}
where, as before, $\theta_e$ corresponds to the dimensionless
electron temperature. Considering the ideal gas equation of state,
one can obtains
\begin{equation}
\label{gas} p=n k_B\left(T_p+T_e\right)=\bar{\rho}(1+z)
\frac{m_e}{m_p} \theta_e\,,
\end{equation}
and substituting Eqs.~\eqref{xi} and~\eqref{gas} into
Eq.~\eqref{energy}, one can derive the following expression after
integration as
\begin{equation}
\theta_e=\left(\theta_e\right)_0\left(\frac{\bar{\rho}}{\bar{\rho}_0}\right)^{\frac{2(1+z)}{3(2+z)}}\,,
\end{equation}
with $(\theta_e)_0=\theta_e(r_0)$. Considering an infalling flow
that satisfies Eq.~\eqref{eq:im}, the analytical expressions for the
rest-mass density and the electron temperature become
\begin{equation}
\begin{aligned}
    \bar{\rho}(r,\theta) &=\bar{\rho}(r_h,\theta) \left(\frac{r_h}{r}\right)^2 \sqrt{\frac{1}{1-g^{rr}}}=\bar{\rho}(r_h,\theta) \sqrt{\frac{r_h^4}{2 r^3 +r}}\,,\\
    \theta_e(r,\theta) &=\theta_e(r_h,\theta)\left(\frac{r_h}{r}\right)^{\frac{4(1+z)}{3(2+z)}} \left(\frac{1}{1-g^{rr}}\right)^{\frac{1+z}{3(2+z)}}=\theta_e(r_h,\theta) \left( \frac{r_h^4}{2 r^3 +r}\right)^{\frac{1+z}{3(2+z)}}\,.
\end{aligned}
\end{equation}

\begin{figure}[H]
    \centering
    \subfigure[$\eta=0.1,\theta_o=0^\circ$]{\includegraphics[scale=0.4]{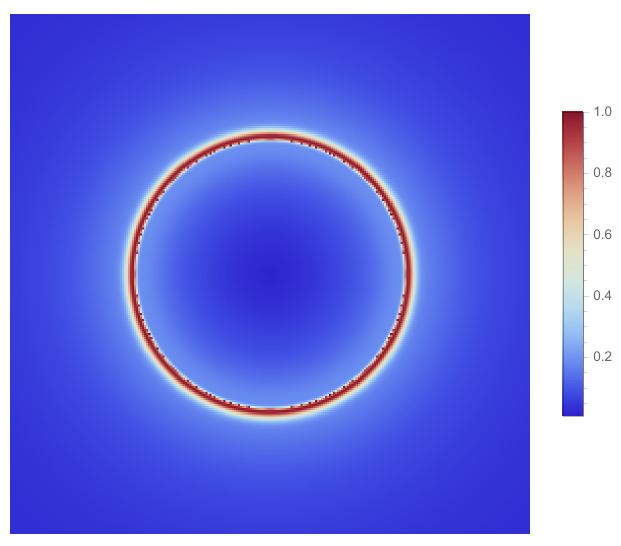}}
    \subfigure[$\eta=0.1,\theta_o=17^\circ$]{\includegraphics[scale=0.4]{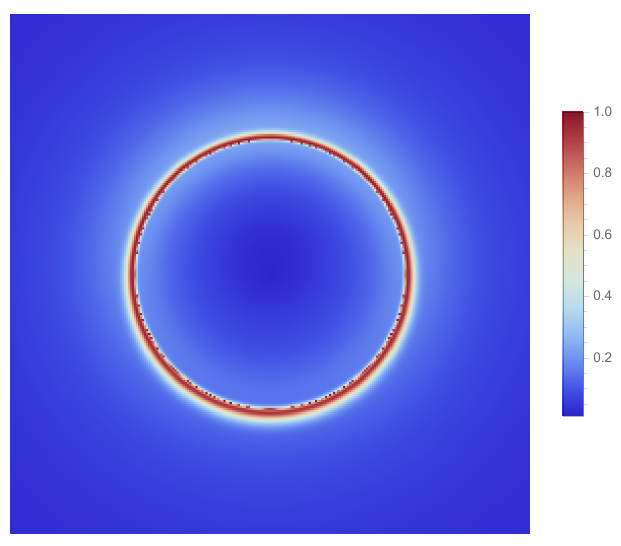}}
    \subfigure[$\eta=0.1,\theta_o=70^\circ$]{\includegraphics[scale=0.4]{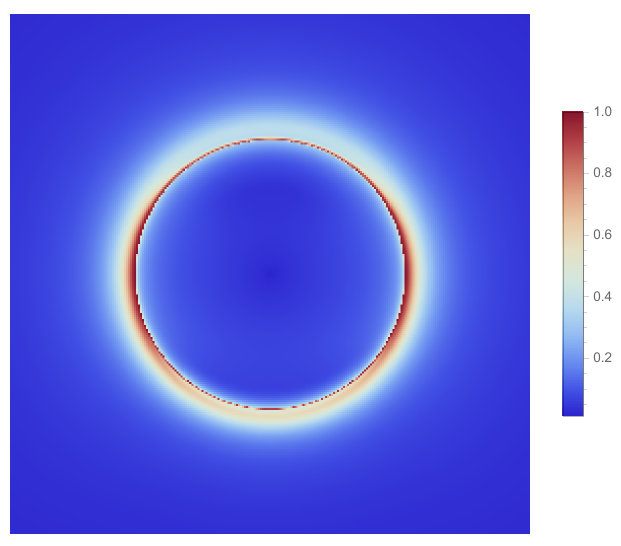}}

    \subfigure[$\eta=0.3,\theta_o=0^\circ$]{\includegraphics[scale=0.4]{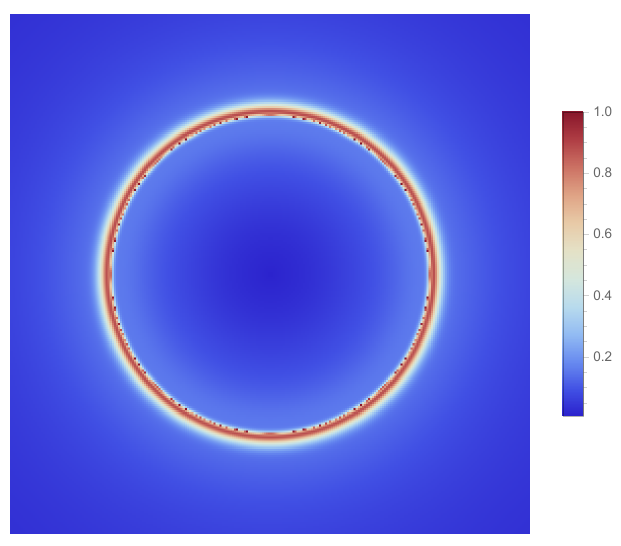}}
    \subfigure[$\eta=0.3,\theta_o=17^\circ$]{\includegraphics[scale=0.4]{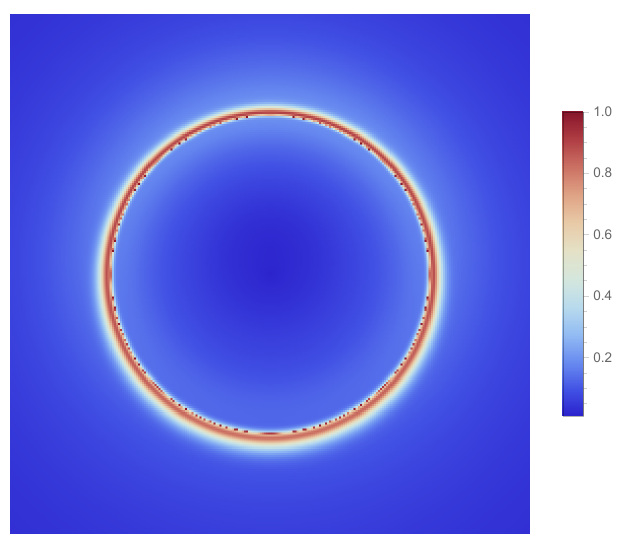}}
    \subfigure[$\eta=0.3,\theta_o=70^\circ$]{\includegraphics[scale=0.4]{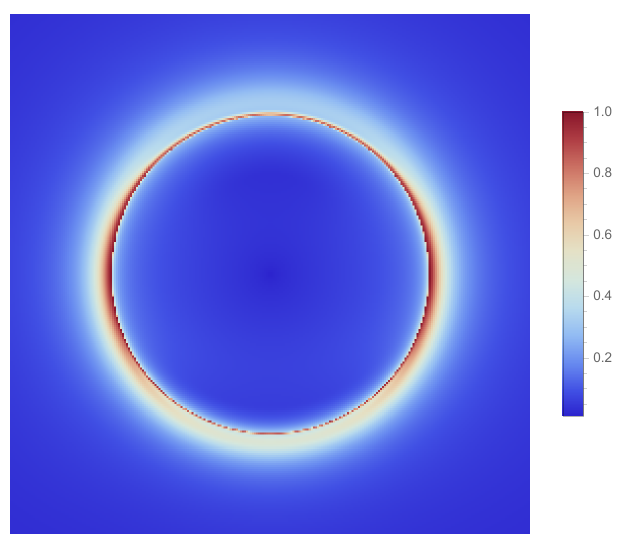}}

        \subfigure[$\eta=0.5,\theta_o=0^\circ$]{\includegraphics[scale=0.4]{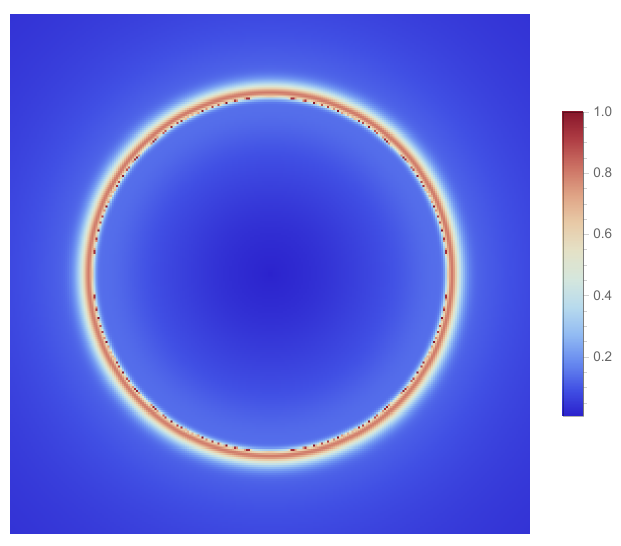}}
    \subfigure[$\eta=0.5,\theta_o=17^\circ$]{\includegraphics[scale=0.4]{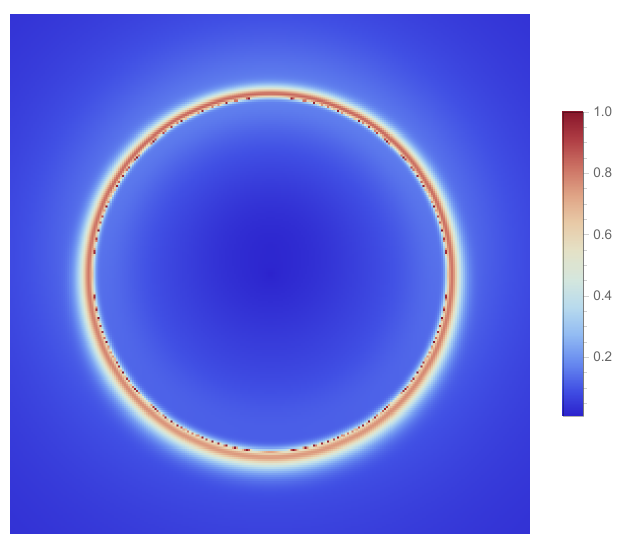}}
    \subfigure[$\eta=0.5,\theta_o=70^\circ$]{\includegraphics[scale=0.4]{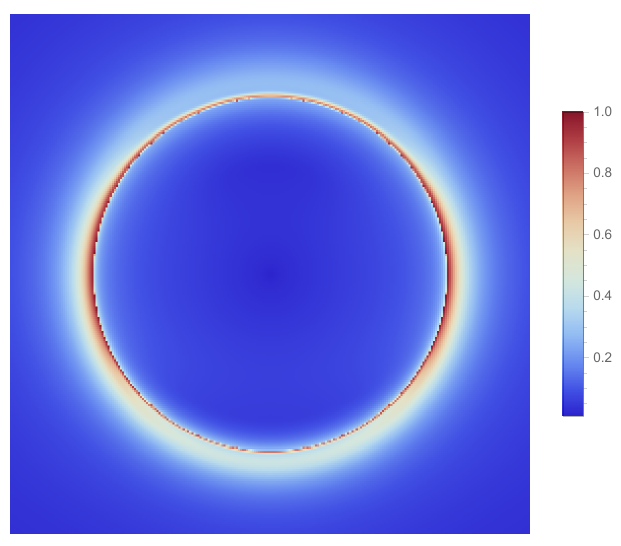}}

        \caption{Effects of $\eta$ and $\theta_o$ on the HOU disk model at the observation frequency of
$230\,\mathrm{GHz}$, with the accretion flow is the infalling
motion.}\label{fig5}
\end{figure}

Subsequently, the angular dependence of $\bar{\rho}(r_h,\theta)$ is
modeled by a Gaussian profile, while in the conical solution, we
choose $\theta_e(r_h,\theta)$ to be constant:
\begin{equation}
\bar{\rho}\left(r_h, \theta\right)=\bar{\rho}_h \exp
\left[-\left(\frac{\sin \theta-\sin
\theta_J}{\bar{\sigma}}\right)^2\right], \quad \theta\left(r_h,
\theta\right)=\theta_h\,.
\end{equation}
Here, $\theta_J$ specifies the central latitude of the distribution
and $\bar{\sigma}$ its angular width. We consider an equatorially
symmetric thick disk with $\theta_J=\pi/2$ and $\bar{\sigma}=1/5$.
For the $\text{M87}^*$ BH, observational estimates suggest $\rho_h
\simeq 1.5 \times 10^3~\mathrm{g\,cm^{-3}\,s^{-2}}$, and $\theta_h
\simeq 16.86$, corresponding to $n_h \simeq 10^6~\mathrm{cm^{-3}}$,
$T_h \simeq 10^{11}~\mathrm{K}$. Assuming stationarity, axisymmetry,
and the ideal MHD condition~\cite{Ruffini}, the magnetic field takes
the general form~\cite{hou2024new,Ruffini}
\begin{equation}
\label{beq} B^\mu=\frac{\Psi}{\sqrt{-g} u^r}\left(\left(u_t+\varpi_B
u_\phi\right) u^\mu+\delta_t^\mu+\varpi_B \delta_\phi^\mu\right)\,,
\end{equation}
where $\Psi=F_{\theta\phi}$ is a component of the electromagnetic
tensor and $\varpi_B$ denote the field line angular velocity, which
characterizes the rotation of magnetic field lines. The spatial
part $B^i$ is parallel to $u^i$, indicating that the magnetic field
is frozen into the streamlines. For simplicity and without loss of
generality, we set $\varpi_B=0$ in what follows. In this work, we
impose a separable magnetic monopole
configuration~\cite{Blandford:1977ds}
\begin{equation}
\Psi=\Psi_0 \operatorname{sign}(\cos \theta) \sin \theta\,.
\end{equation}
Finally, based on the aforementioned mechanism, we present the BH
images illuminated by the BAAF disk and investigate their
corresponding intensity and polarization distributions. In Fig.
\textbf{\ref{fig5}}, we illustrate the intensity maps of the
Schwarzschild BH surrounded by PFDM within the mechanism of Hou
disk model and the accretion flow is the infalling motion with fixed
observed frequency is $230\,\mathrm{GHz}$. Each row and column
corresponds to a particular value of the PFDM parameter $\eta$ and
observer's inclination $\theta_o$, as indicated. In all images, the
bright ring that lies in the center of the screen originates from higher
order images, while the central dark region corresponds to the BH
event horizon. Overall, the impact of relevant parameters such as
$\eta$ and observer's inclination $\theta_o$, showing the similar
behavior, as observed in the RIAF thick disk model. However, a few
key distinctions emerge. Compared with the RIAF thick disk model,
the bright photon ring in the Hou disk framework displays generally
thinner, and the gap between the primary and higher-order images
becomes more obvious. Additionally, at higher values of observers
positions $\theta_o$, the higher-order image does not interpret the
two distinct dark interior, as observed in the RIAF-like model. This
indicates that, in the RIAF model, radiation from off-equatorial
regions more effectively obscures the event-horizon silhouette. Such
differences may arise because, for certain parameter choices, the
Hou disk modeled under the conical approximation becomes
geometrically thinner than the corresponding RIAF-like disk model in
specific regions.

\begin{figure}[H]
    \centering
    \subfigure[Stokes parameter $\mathcal{I}_{o}$]{\includegraphics[scale=0.5]{Po1_1}}
    \subfigure[Stokes parameter $\mathcal{Q}_{o}$]{\includegraphics[scale=0.6]{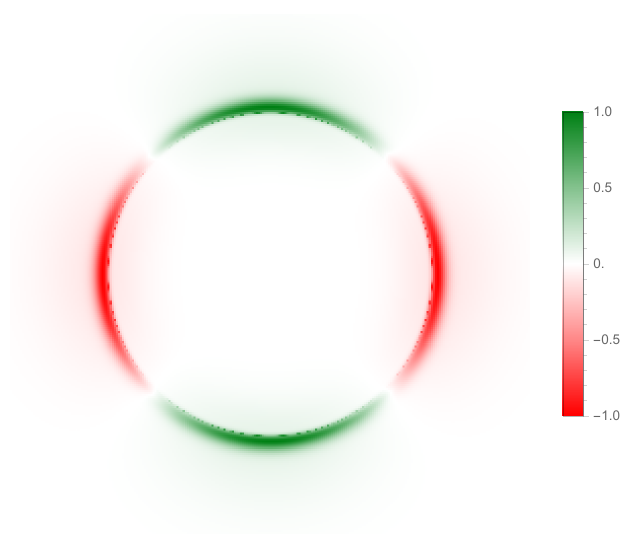}}
    \subfigure[Stokes parameter $\mathcal{U}_{o}$]{\includegraphics[scale=0.6]{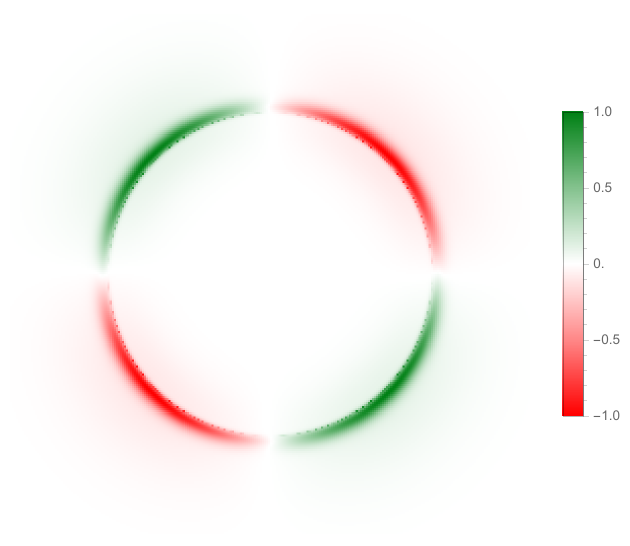}}
    \subfigure[Stokes parameter $\mathcal{V}_{o}$]{\includegraphics[scale=0.6]{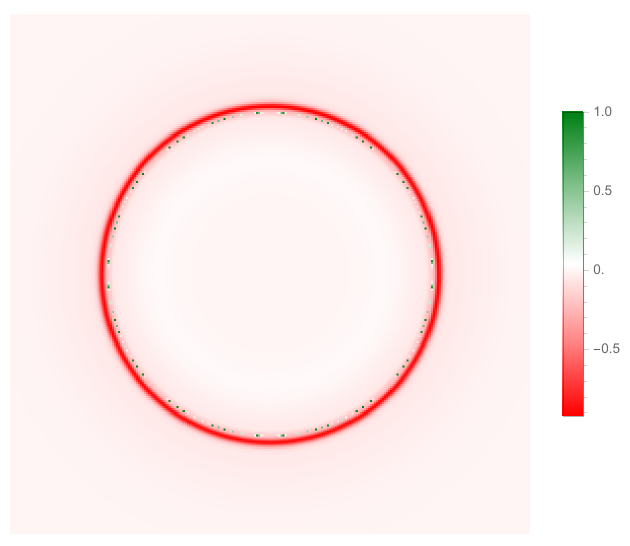}}

        \caption{Stokes parameters for $\eta=0.1$, $\theta_o=0^\circ$, at $230\,\mathrm{GHz}$ with the accretion flow is the infalling motion.}\label{fig6}
\end{figure}

Now we investigate the polarization properties, as presented in Fig.
\textbf{\ref{fig6}}, where we observed that the spatial
distributions of the observed Stokes parameters, $\mathcal{I}_{o},
\mathcal{Q}_{o}, \mathcal{U}_{o}, \mathcal{V}_{o}$ for the
Schwarzschild BH surrounded by PFDM in the Hou disk model. In this
case, the accretion flow motion mode is the infalling motion, with
fixed parameters $\eta=0.1$, and $\theta_o=70^\circ$, and the
observing frequency is $230\,\mathrm{GHz}$. The $\mathcal{I}_{o}$
panel indicates the overall intensity distribution, where the arrows
represent the electric vector position angle (EVPA),
$\Phi_{\mathrm{EVPA}}$, and the color corresponds to the linear
polarization degree, $P_{o}$. Since the EVPA is always perpendicular
to the global magnetic field, $B_{\mu}$, the polarization pattern
suggests that the magnetic field is approximately radial. The
combined distributions of $\mathcal{Q}_{o}$ and $\mathcal{U}_{o}$
qualitatively evaluate the EVPA direction, while $\mathcal{V}_{o}<0$
represents the right-handed polarization.

\begin{figure}[H]
    \centering
    \subfigure[$\eta=0.1,\theta_o=0^\circ$]{\includegraphics[scale=0.4]{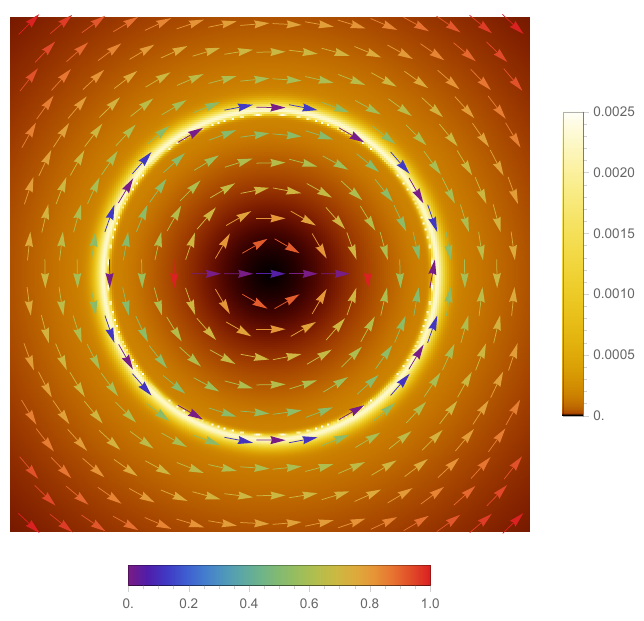}}
    \subfigure[$\eta=0.1,\theta_o=17^\circ$]{\includegraphics[scale=0.4]{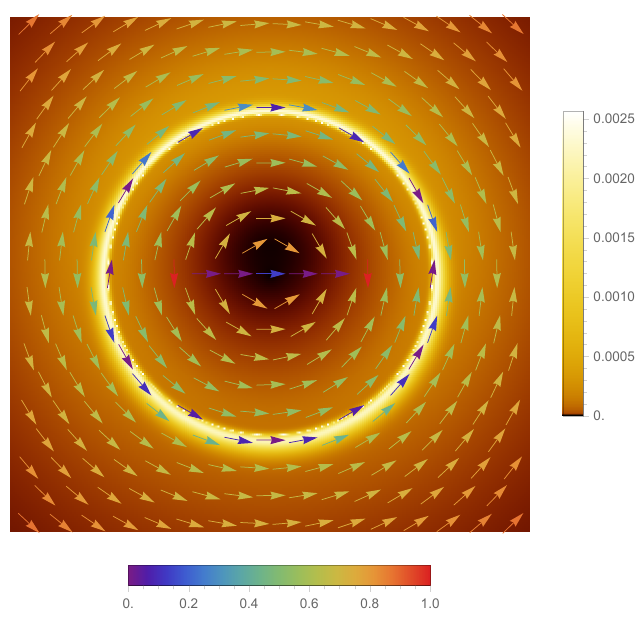}}
    \subfigure[$\eta=0.1,\theta_o=70^\circ$]{\includegraphics[scale=0.4]{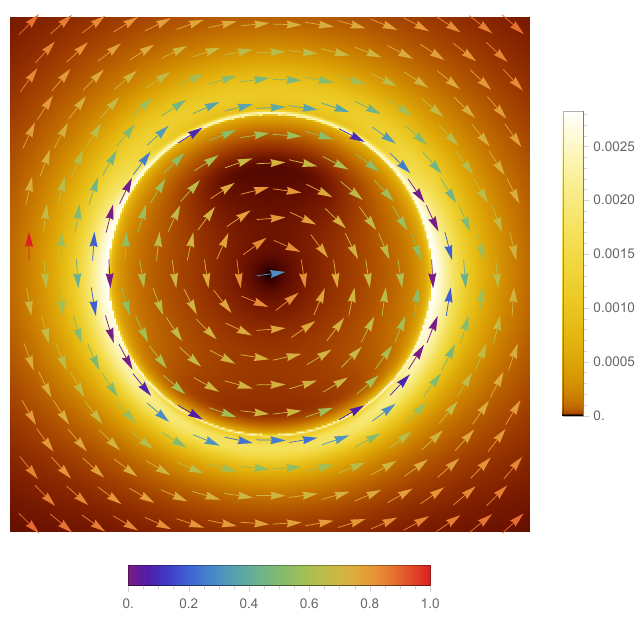}}

    \subfigure[$\eta=0.3,\theta_o=0^\circ$]{\includegraphics[scale=0.4]{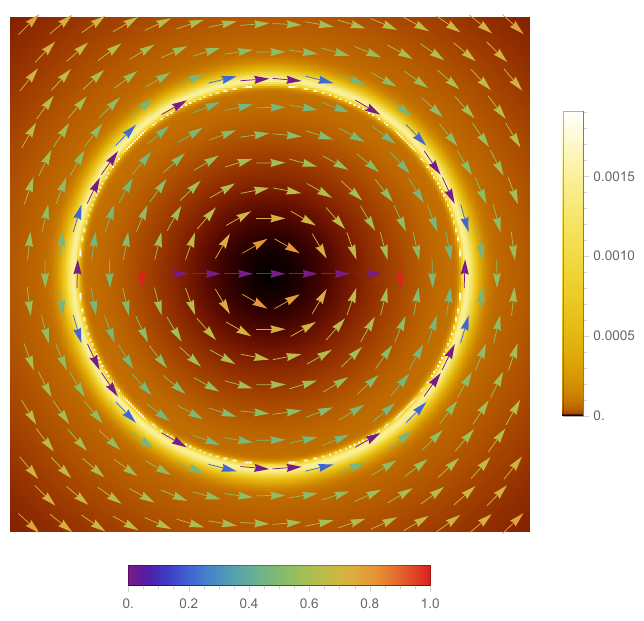}}
    \subfigure[$\eta=0.3,\theta_o=17^\circ$]{\includegraphics[scale=0.4]{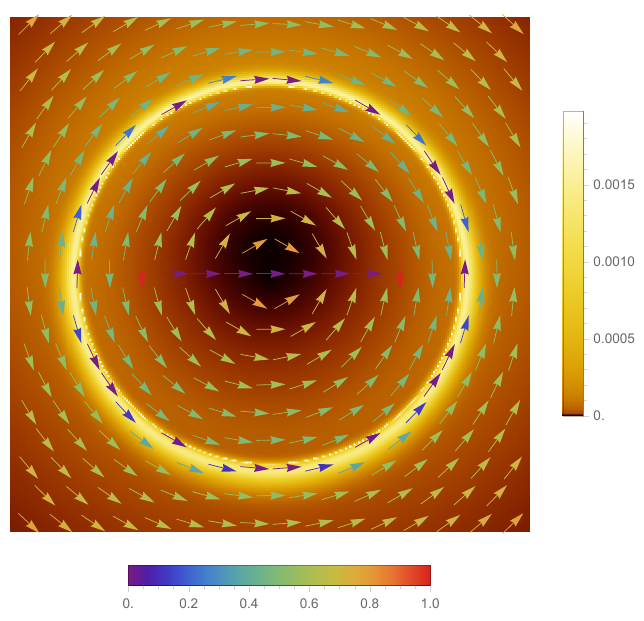}}
    \subfigure[$\eta=0.3,\theta_o=70^\circ$]{\includegraphics[scale=0.4]{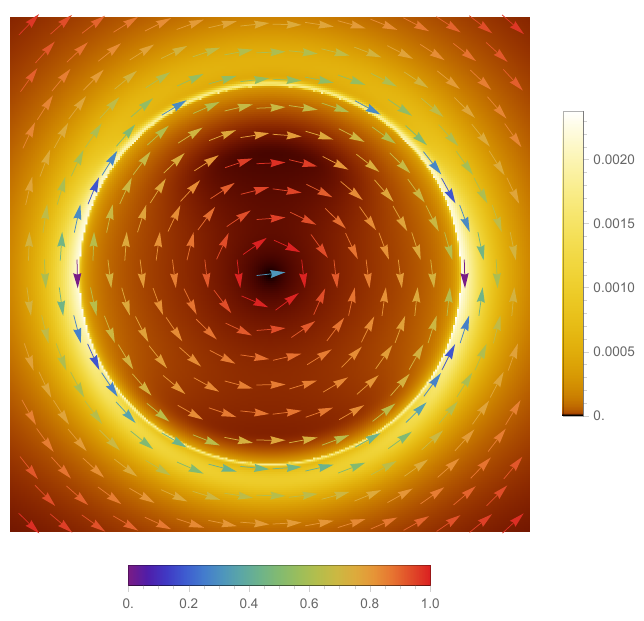}}

        \subfigure[$\eta=0.5,\theta_o=0^\circ$]{\includegraphics[scale=0.4]{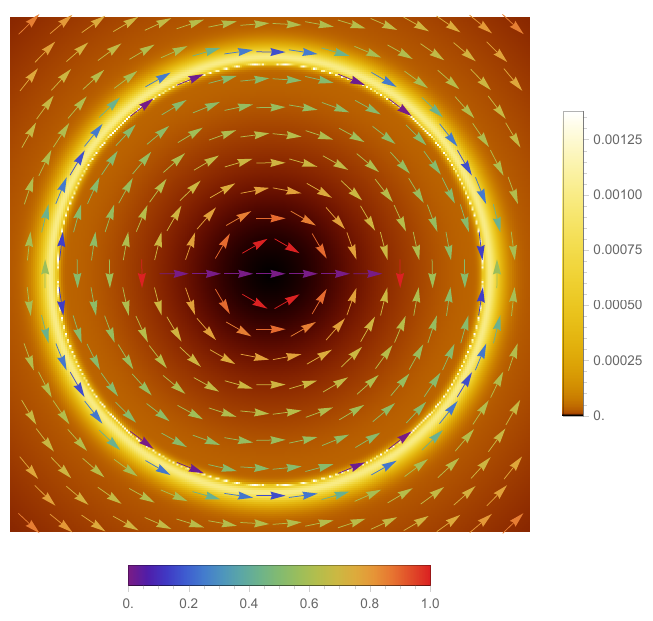}}
    \subfigure[$\eta=0.5,\theta_o=17^\circ$]{\includegraphics[scale=0.4]{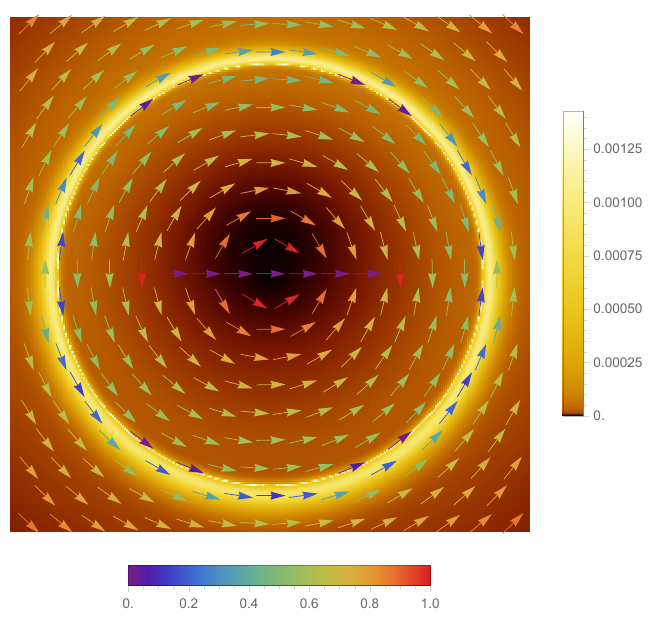}}
    \subfigure[$\eta=0.5,\theta_o=70^\circ$]{\includegraphics[scale=0.4]{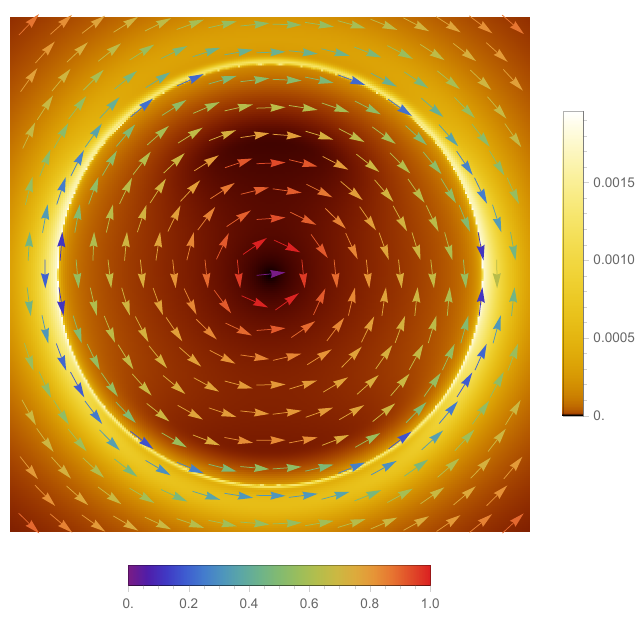}}

    \caption{Polarization image of the HOU disk model at $230\,\mathrm{GHz}$ with the accretion flow is the infalling motion.}\label{fig7}
\end{figure}

Figure \textbf{\ref{fig7}}, exhibit the polarized images of the
Schwarzschild BH surrounded by PFDM in the Hou disk model, for
different values of the PFDM parameter $\eta$ and the inclination
angle $\theta_o$. Moreover, the accretion flow mode is infalling
motion, and the observed frequency is $230\,\mathrm{GHz}$. The
polarization patterns show a pronounced dependence on relevant
parameters, with notable variations in morphology and strength
of the polarized flux across the different panels. Overall, the EVPA
pattern remains closely perpendicular to the radial direction,
consistent with the assumed alignment of the magnetic field along
the radially infalling motion. In the surroundings of the
higher-order images, the polarization displays rapid spatial
variations. Here, we observed that as the values of $\eta$
increases from top to bottom, both the size of the higher-order
images as well as the central dark region are enhanced
significantly. Moreover, the corresponding direction of EVPA
distribution has changed. Across each row, the increasing values of
$\theta_o$ results in slightly deviating the EVPA from the purely
azimuthal orientation, while the polarized intensity within the dark
region becomes more prominent. It is worth noting that, in thin
disk models, the inner shadow corresponds to the BH horizon, where
no polarization is present in this region. On the contrary, in the present
scenario, for geometrically thick disks, gravitational lensing
allows emission from regions above and below the equatorial plane to
overlap with the projected horizon boundary, producing
polarization vectors that distribute the entire image plane.

\begin{figure}[H]
       \centering

       \subfigure[Horizontal, HOU disk]{\includegraphics[scale=0.8]{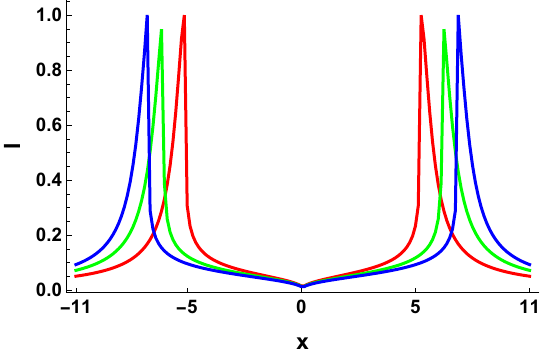}}
       \subfigure[Vertical, HOU disk]{\includegraphics[scale=0.8]{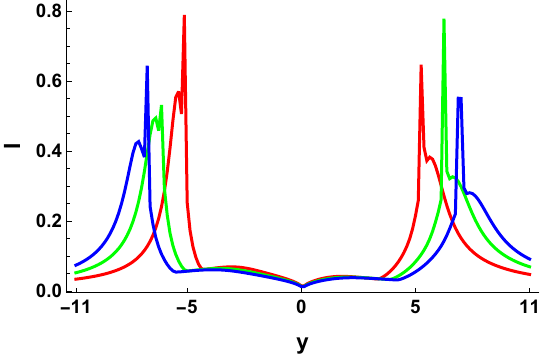}}

        \caption{The intensity cuts with $\theta_o=70^\circ$ and the accretion flow is the infalling motion. The red, green, and blue curves correspond to $\eta=0.1,\,0.3,\,0.5$, respectively.}\label{fig8}
\end{figure}

In Fig. \textbf{\ref{fig8}}, we quantitatively investigate the
horizontal and vertical intensity cuts for the BAAF model with fixed
$\theta_o=70^\circ$ and for different values of $\eta$ at observed
frequency $230\,\mathrm{GHz}$. In both panels, the peak of the curves
corresponds to the higher-order images in the BAAF model, which
significantly varies with the variation of $\eta$, and the intensity
rapidly approaches zero beyond the peaks. These results show that,
in the BAAF model, the intensity distribution originally comes from
the higher-order images. On the other hand, in the RIAF model, the
intensity of the primary image beyond the higher-order images
remains relatively greater.

\section{Conclusion}
In this work, we considered the Schwarzschild BH in the presence of
PFDM and investigate the impact of relevant parameters on the visual
characteristics of the geometrically thick accretion flows. We first
review the background of the BH model, geodesic equations for the
radius of the photon sphere, as well as the ZAMO frame, to observe the
corresponding images on the screen. We then considered two models of
geometrically thick accretion flows, such as a phenomenological
RIAF-like thick disk model and an analytical BAAF model. To obtain
the desired results, we numerically solve the null geodesics and
radiative transfer equations, and investigate the intricate
properties of thermal synchrotron radiation in the magnetofluid
through effective implementations.

In the case of the RIAF-like model, we first discussed the isotropic
emission scenario along with the infalling accretion flow with fixed
observed frequency at $230\,\mathrm{GHz}$. The obtained results
show the increasing values of $\eta$ from top to bottom, resulting in
an increase in both the bright ring and the central dark region without
altering their shapes. The variation in inclination angle
$\theta_{o}$ shows that for polar observers, the brightness
distribution is nearly axisymmetric. However, for higher inclination
angles, the brightness becomes asymmetric, and the central dark
region splits into two parts due to off-equatorial emission.
Additionally, the width of the bright ring is slightly enhanced on
the left and right sides of the screen with the aid of $\theta_{o}$.
Subsequently, we investigated the anisotropic synchrotron emission
by imposing a toroidal magnetic field and infalling motion.
Compared with the isotropic scenario, the corresponding brightness
distribution becomes appreciable asymmetric for higher values of
inclination angles. The bright photon ring displays vertically
elongated and slightly elliptical, depicting the geometry of the
underlying magnetic field configuration.

Finally, in the case of the BAAF model, we observed that the distribution
of higher-order images is narrower compared to the RIAF model, and
the influence of radiation from outside of the equatorial plane on
the inner shadow is appreciably reduced. This difference originates
due to the choice of some particular parameter values; the Hou disk
in the conical approximation is geometrically thinner as compared to
RIAF-like disk in some regions. For the polarized images, the
polarization intensity closely tracks the overall brightness
distribution, with progressing polarization occurring in the
brighter regions. Both the polarization degree and the EVPA
orientation interpret a clear dependence on the PFDM parameter as
well as the inclination angle, indicating that the polarization
signatures of a Schwarzschild BH with PFDM provide a promising probe
of the underlying spacetime geometry. Finally, we summarize this
manuscript with various future perspectives. In the near-horizon
polarization, we have constrained particular parameter choices of
BAAF model, like as disk thickness and temperature. Although the
underlying BH model may exhibit intrinsic near-horizon polarization,
a variety of astrophysical effects can obscure these theoretical
characteristics. Hence, separating the gravitational effects from
those arising due to plasma and radiative processes remains a
crucial yet challenging task for future investigations.\\\\
{\bf Acknowledgements}\\
This work is supported by the National Natural Science Foundation of
China (Grants Nos. 12375043, 12575069 ).

\end{document}